\documentclass[oneside]{aa}

\usepackage[varg]{txfonts}
\usepackage{natbib}
\usepackage{graphicx}
\usepackage{subcaption}
\usepackage{wasysym} 
\usepackage{rotating}
\usepackage{multirow} 
\usepackage{xcolor} 
\usepackage{amsmath}
\usepackage{ulem} 
\usepackage{blindtext}
\usepackage{color}
\usepackage[title,toc]{appendix}
\usepackage{threeparttable}
\usepackage{amssymb}
\usepackage{longtable}
\graphicspath{{Figure/}}
\usepackage{booktabs}
\usepackage{pdflscape}
\usepackage{placeins}

\DeclareMathAlphabet{\mathsc}{OT1}{cmr}{m}{sc}
\def\testbx{bx}
\DeclareRobustCommand{\ion}[2]{
\relax\ifmmode
\ifx\testbx\f@series
{\mathbf{#1\,\mathsc{#2}}}\else
{\it{#1\,\mathsc{#2}}}\fi
\else\textup{#1\,{\mdseries\textsc{#2}}}
\fi}

\begin{document}

\title{Supernovae at Distances $<$ 40 Mpc: II.}
\subtitle{Supernova Rate in the Local Universe}

\author{
Xiaoran Ma \inst{\ref{inst1}} 
\and Xiaofeng Wang \inst{\ref{inst1},\ref{inst2} \thanks{E-mail: wang\_xf@mail.tsinghua.edu.cn}}
\and Jun Mo \inst{\ref{inst1}}
\and D. Andrew Howell \inst{\ref{inst3},\ref{inst4}}
\and Craig Pellegrino \inst{\ref{inst3},\ref{inst4}}
\and Jujia Zhang \inst{\ref{inst5},\ref{inst6}}
\and Chengyuan Wu \inst{\ref{inst5},\ref{inst6}}
\and Shengyu Yan \inst{\ref{inst1}}
\and Dongdong Liu \inst{\ref{inst5},\ref{inst6}}
\and Iair Arcavi \inst{\ref{inst7}}
\and Zhihao Chen \inst{\ref{inst1}}
\and Joseph Farah \inst{\ref{inst3},\ref{inst4}}
\and Estefania Padilla Gonzalez \inst{\ref{inst3},\ref{inst4}}
\and Fangzhou Guo \inst{\ref{inst1}}
\and Daichi Hiramatsu \inst{\ref{inst8},\ref{inst9}}
\and Gaici Li \inst{\ref{inst1}}
\and Han Lin \inst{\ref{inst5},\ref{inst6}}
\and Jialian Liu \inst{\ref{inst1}}
\and Curtis McCully \inst{\ref{inst3},\ref{inst4}}
\and Megan Newsome \inst{\ref{inst3},\ref{inst4}}
\and Hanna Sai \inst{\ref{inst1}}
\and Giacomo Terreran \inst{\ref{inst3},\ref{inst4}}
\and Danfeng Xiang \inst{\ref{inst1}}
\and Xinhan Zhang \inst{\ref{inst10}}
}

\institute{
Department of Physics, Tsinghua University, Haidian District, Beijing 100084, China \label{inst1}
\and Purple Mountain Observatory, Chinese Academy of Sciences, Nanjing 210023, China  \label{inst2}
\and Las Cumbres Observatory, 6740 Cortona Drive Suite 102, Goleta, CA 93117-5575, USA \label{inst3}
\and Department of Physics, University of California, Santa Barbara, CA 93106-9530, USA \label{inst4}
\and Yunnan Observatories, Chinese Academy of Sciences, Kunming 650216, China \label{inst5}
\and International Centre of Supernovae, Yunnan Key Laboratory, Kunming 650216, P. R. China \label{inst6}
\and School of Physics and Astronomy, Tel Aviv University, Tel Aviv 69978, Israel \label{inst7}
\and Center for Astrophysics \textbar{} Harvard \& Smithsonian, 60 Garden Street, Cambridge, MA 02138-1516, USA \label{inst8}
\and The NSF AI Institute for Artificial Intelligence and Fundamental Interactions, USA \label{inst9}
\and School of Physics and Information Engineering, Jiangsu Second Normal University, Nanjing 211200, China \label{inst10}
}

\date{Received Month XX, 2022 / Accepted Month XX, 2022}

\abstract{This is the second paper of a series aiming to determine the birth rates of supernovae (SNe) in the local Universe.}{In this paper, we aim to estimate the SN rates in the local universe and fit the delay-time distribution of SNe Ia to put constraints on their progenitor scenarios.}{We performed a Monte-Carlo simulation to estimate the volumetric rates with the nearby SN sample introduced in Paper I of the series. The rate evolution of core-collapse (CC) SNe well traces the evolution of cosmic star formation history; while the rate evolution of SNe Ia involves the convolution of cosmic star-formation history and a two-component delay-time distribution including a power law and a Gaussian component.} {
The volumetric rates of type Ia, Ibc and II SNe are derived as $0.325\pm0.040^{+0.016}_{-0.010}$, $0.160\pm0.028^{+0.044}_{-0.014}$, and $0.528\pm0.051^{+0.162}_{-0.013}$ (in unit of $10^{-4} \mathrm{yr^{-1} Mpc^{-3} h^3_{70}}$), respectively. The rate of CCSNe ($0.688\pm0.078^{+0.206}_{-0.027}$) is consistent with previous estimates, which traces the star-formation history. On the other hand, the newly derived local SN Ia rate is larger than existing results given at redshifts 0.01 < z < 0.1, favoring for an increased rate from the universe at z $\sim$ 0.1 to the local universe at z $<$ 0.01. A two-component model can well fit the rate variation, with the power law component accounting for the rate evolution at larger redshifts and the Gaussian component with a delay time of 12.63$\pm$0.38 Gyr accounting for the local rate evolution, respectively. This latter delayed component with such a longer delay time suggests that the progenitors of these SNe Ia were formed at around 1 Gyr after the birth of the universe, which could only be explained by a double-degenerate progenitor scenario. This is evidenced by the comparison with the PTF sample of SNe Ia at z = 0.073 and the morphology of their host galaxies, which reveals that the increase in SN Ia rate at z $<$ 0.01 is primarily due to the SNe Ia of massive E and S0 galaxies with old stellar populations. Based on the above results, we estimate the Galactic SN rate as 3.08$\pm$1.29 per century.  
}
{}
\authorrunning{Ma et al.} 
\titlerunning{Supernova Rate in the Local Universe}
\keywords{supernovae: general – methods: data analysis – surveys}
\maketitle

\section{Introduction}

The birth rates of different types of SNe and their evolution with respect to the redshift provide important constraints on SN progenitors and advance our understanding of cosmic chemical evolution.
In the $20^{th}$ century and the first decade of the $21^{st}$ century, many studies have attempted to measure SN rates in local and distant universe based primarily on targeted surveys of preselected galaxies or fields of the sky. 
The conventional procedure is the control-time method \citep{Zwicky1942, van1991, Leaman2011}, which involves the construction of light curve functions for different types of SNe. For each type, the sum of time when the SNe would be brighter than the limiting magnitude of the survey is defined as the "control time" of the search. The SN rate is then calculated using the total number of discovered samples divided by the control time. It is usually expressed in units of SNu, for instance $1~\mathrm{SNuB} = 1~\mathrm{SN (100 yr)^{-1} (10^{10} L^B_{\odot})^{-1}}$, where $\mathrm{L^B_{\odot}}$ represents the B-band solar luminosity. With the galaxy luminosity distribution, the SN rate expressed in units of SNu can be converted into a volumetric rate. However, the sample from target surveys might introduce an observational bias, due to a preference of monitoring brighter galaxies.

Before the 1990s, the sample used for the studies of SN rates came from the Palomar SN search \citep{Zwicky1942}, the Asiago SN search \citep{Cappellaro1988}, and Robert Evans’ visual search \citep{van1987, van1990}. After that, \citet{Cappellaro1999} collected an important sample of 137 SNe from historical SN surveys, which became the most valuable one in studies of SN rates, host galaxy environments, SN progenitor systems, and cosmic star-formation history \citep{Mannucci2005, Mannucci2005B, Mannucci2008}.
Later, \citet{Li2011b} established a complete sample of 175 SNe and a “full-optimal” SN sample with a total of 726 SNe from the 10-year Lick Observatory Supernova Search (LOSS) program. With this homogeneous set of nearby SNe from a single survey, they derived the most accurate estimates of the fractions of different types of SNe and the corresponding rates in the local universe at that time. Using data from the Palomar Transient Facility \citep[PTF,][]{Rau2009, Law2009}, \citet{Frohmaier2019, Frohmaier2021} gave one of the most updated rate measurements for type Ia and CC SNe at z $<$ 0.1.

SN rates at moderate to higher redshifts have been derived with SN samples from untargeted rolling searches \citep{Dahlen2004, Dilday2010, Perrett2012, Rodney2014, Cappellaro2015, Frohmaier2019, Frohmaier2021}, which help better constrain rate evolution and progenitor systems of different types of SNe. In particular, the evolution of CCSN rates with redshifts is found to track well the cosmic star formation history \citep[SFH,][]{Hopkins2006, Rujopakarn2010, Cucciati2012, Frohmaier2021}, as CCSNe usually originate from massive stars with short lifetimes.  

In comparison, the evolution of the SN Ia rate does not track the SFH but can be regarded as the convolution of delay-time distribution (DTD) and SFH. The delay time is defined as the duration between the instantaneous burst of star formation and the final resulting SN Ia explosions. Among the different parameterization models of DTD, a simple power-law model, that is, $\propto t^{-\beta}$ with $\beta \simeq 1$, turns out to be valid \citep{Maoz2012b, Graur2013, Graur2014}. Other models include the functional form of e-folding, Gaussian \citep{Dahlen2012,Palicio2024}, and the two-component model. A popular form of the two-component model is comprised of a prompt component that tracks the instantaneous star-formation rate (SFR) and a delayed component that is proportional to stellar mass \citep{Mannucci2005}. The prompt component represents very young SNe Ia that explode soon after the formation of their progenitors, while the delayed component has longer delay times and corresponds to older stellar population. With reliable measurements of SN Ia rate with redshift and cosmic SFH, the DTD can be determined by inverting the convolution \citep{Horiuchi2010, Dahlen2012, Perrett2012, Graur2014, Rodney2014, Frohmaier2019}. Different DTDs can give insight to the progenitor systems of SNe Ia \citep{Maoz2012}. For example, the DD channel with two CO WDs can provide a DTD for normal SNe Ia with an initial peak at around 1 Gyr and a tail up to about 10 Gyr \citep{Pakmor2013}, while most SD models give short delay times with few or no SNe Ia produced beyond 2-3 Gyr \citep{Childress2014,Maoz2014}.

Paper I of this series discusses the construction of the nearby SN and galaxy samples. A total of 211 SNe discovered over the years from 2016 to 2023 are selected with a distance cut at 40 Mpc. The SN sample consists of 69 SNe Ia, 34 SNe Ibc and 109 SNe II. In this paper, two galaxy samples are used, including the host galaxy sample with a total of 191 galaxies and the GLADE+ sample with a total of 8790 galaxies. The Hubble-type distributions of the two galaxy samples show noticeable differences. The most abundant type in the local universe is the E, S0, Scd and Irr galaxies, whereas the galaxies hosting most SNe are of type Sc. The average stellar mass distribution suggests that the galaxies hosting SNe generally tend to be more massive. For all of our SN sample, we obtained their classifications and gave detailed subtype fractions. Then, combined with host galaxy information, we studied the radial and stellar mass distributions of different subtypes and their correlations. 
The number distribution of SNe in galaxies of different Hubble types is compared to that of the SN sample from \citet{Li2011}. We find clear evidence of a double-peak structure in E-S0 and late-type Sc galaxies for the SN Ia sample. This could suggest a two-component model for SN Ia DTD, with a prompt and a delayed component corresponding to the young and old stellar population in late-type spirals and E-S0 galaxies, respectively.

This is paper II of the series and is organized as follows. In Sect.~\ref{Rate}, we describe the method of estimating local volumetric rates and the final results. We calculate the SN rate in galaxies of different Hubble types in Sect.~\ref{SNuM}. Then we make use of our results to give the SN rate of the Milky Way. In Sect.~\ref{analysis}, we compare our rate measurements with historical results and derive the DTDs for SNe Ia using both our new result and those of the literature. The CCSNe rate evolution is fitted with the cosmic star-formation history. We summarize in Sect.~\ref{sum}.
	
\section{The volumetric supernova rate} \label{Rate}
	
In this section, we describe our approach to estimating the volumetric SN rates using the SN sample established in Paper I and discuss their uncertainties.
	
\subsection{Supernova rate estimation}
	
Following the approach proposed by \citet{Rodney2014}, we define $\mathrm{N_{obs}}$, the observed count, as the observed number of SNe and the control count, $\mathrm{N_{ctrl}}$, as the expected number of SNe that should be detected if the local SN rates were constant at $10^{-4} \mathrm{yr^{-1} Mpc^{-3} h^3_{70}}$. Then the local volumetric SN rate in unit of $10^{-4} \mathrm{yr^{-1} Mpc^{-3} h^3_{70}}$ is given by 

\begin{equation}
    \mathrm{SNR = \frac{N_{obs}}{N_{ctrl}}}.
\end{equation}
	
The observed count, $\mathrm{N_{obs}}$, is simply the number of SNe detected within 40 Mpc between 2016 and 2023. 
The control count, $\mathrm{N_{ctrl}}$, was calculated using Monte Carlo simulations. For each type, Ia, Ibc and II, we generated 100,000 SNe as the total sample. We randomly generated the following set of properties for each individual SN: distance, absolute peak magnitude, host extinction, Galactic extinction, the right ascension, declination, and date of peak brightness.
	
We divided each sample into different subtypes by random sampling according to the fractions given in Paper I. All SNe were randomly located in a sphere with a radius of 40 Mpc under uniform distribution. Then for each subtype, we generated the absolute peak magnitudes based on the bias-corrected Gaussian distributions of B-band peak absolute magnitudes of different subtypes given by \citet{Richardson2014} (see Table ~\ref{MB} and their Table 1 for detailed parameters for the Gaussian distributions). According to the subtype fractions estimated in Paper I, we adopted the brightest 5.9\% (i.e., 91T-like events take up 5.9\% of all SNe Ia) of the generated Ia sample as the 91T sample and the dimmest 17.6\% as the 91bg sample, the rest is taken as the normal Ia sample. Since \citet{Richardson2014} did not give the absolute peak magnitude distribution for the 02cx-like events, we adopted a uniform distribution between $-14.0$ and $-18.0$ mag for this subtype according to the peak magnitude range given by \citet{Jha2017}. The host extinction information was generated based on the results given by \citet{Holwerda2015}, for which the distribution of $\mathrm{A_V}$ is taken as $\mathrm{N = N_0 exp(-A_V/0.4)}$. For the Galactic extinctions, we used the value at the randomly generated SN coordinates (i.e., a uniform distribution in all sky) according to \citet{Schlafly2011}. The date of peak was drawn from a uniform distribution within the range 2016-2023. From the generated properties (i.e., the peak absolute magnitude, distance, host and Galactic extinction value) we can calculate the observed peak apparent magnitude for each SN under extinction effect, and then proceed to determine whether this SN could be detected.

In Paper I we have concluded that the main discoverers of our SN sample were the All-Sky Automated Survey for Supernovae \citep[ASAS-SN,][]{Shappee2014, Kochanek2017}, the Asteroid Terrestrial-impact Last Alert System \citep[ATLAS,][]{Tonry2018, Smith2020}, and the Zwicky Transient Facility \citep[ZTF,][]{Masci2019, Bellm2019}, so we chose to determine whether the generated SNe could be detected by these three surveys. Combining the generated peak apparent magnitudes and the light-curve functions given by \citet{Li2011} (see Table 2 and Figs. 1-3 of \citet{Li2011} for detailed light curve templates). They provided 22 templates for SNe Ia, one each for the 91bg, 91T, and 02cx-like subtypes and 19 for normal Ia; 3 templates for the fast, slow, and average-evolving SNe Ibc and 1 for peculiar SNe Ibc. A single light curve template was constructed for the subtypes of SNe IIP, IIL, and IIb, while three, including fast, average, and slow-evolving, were for SNe IIn. We could simulate the light-curve evolution for the generated SNe and then determine whether it could be actually detected according to their survey strategies \footnote{ASAS-SN: automatically surveying the entire visible sky every night down to about 18 mag, more details can be seen in \citet{Shappee2014} and \citet{Kochanek2017}. 

ATLAS covers about 24,500 deg$^2$ of the sky in the declination range $-45^{\circ}$ < $\delta$ < +90$^{\circ}$ with a cadence of 2 days, with four exposures (over a 1 hour interval) reaching $\sim$ 19.5 mag in the $o$ band when the sky is dark and seeing is good (see details in \citet{Tonry2018} and \citet{Smith2020}).

ZTF scans the entire northern visible sky ($\delta \geqslant -31^{\circ}$ and $\left|b\right| > 7^{\circ}$) every three nights (since the end of 2020, the ZTF public survey has increased its observing cadence to 2 days) at a rate of $\sim$3760 deg$^2$/hour to median depths of g $\sim$ 20.8 and r $\sim$ 20.6 mag, see \citet{Masci2019} and \citet{Bellm2019} for more details. We particularly notice that the ZTF scheduling algorithm is publicly available under an open source license: \url{https://github.com/ZwickyTransientFacility/ztf_sim}.}. For each SN, the light-curve function is randomly chosen from the light curve families of \citet{Li2011} of the given subtype. 

Since the photometric bands of the generated peak absolute magnitudes (B band), light curve templates (R-band), and surveys (see Appendix B) are all different, we need to standardize the peak absolute magnitudes to R band to simulate the evolution of the light curve and then transform them to the corresponding photometric band of the survey to determine the detection probability. To do this, we need statistical patterns of the differences in magnitudes between those photometric bands. For SNe Ia, we follow the work of \citet{Nugent2002}\footnote{Photometric data in UBVRI bands available in \url{https://c3.lbl.gov/nugent/nugent_templates.html}} who presented the UBVRI magnitude differences from the peak B-band value for different subtypes of SNe Ia from 20 days before to 70 days after the B-band maximum. By stretching the templates (between 0.8 < s < 1.1), we could estimate the systematic uncertainties of adopting these templates. For CCSNe, we adopted the work of \citet{Pessi2023}. They provided photometric data in B-, V-, and R-band of CCSNe with detailed subtype classifications, including Ib, Ic, Ic-BL, IIP, IIL, IIb, and IIn (see detailed information in Appendix G of their paper). The corresponding uncertainties were estimated from the errors quoted for the peak magnitudes and post-peak decline rates. Thus, the overall pattern for this procedure is that we randomly generate the B-band peak absolute magnitude for an SN. Then we convert the generated B-band peak magnitudes into the R-band values to fit the light curve templates of \citet{Li2011}. Finally, the generated B-band magnitudes are converted to the values of the corresponding photometric band of the survey to compare with the limiting magnitudes. For the transformation between R- and o-band magnitudes, we adopt the same method as described in \citet{XiangD2019} (see their Section 3.2 and Figure 7 for details of the evolution of c-V and o-R colors with respect to evolution phases.)  

We divided the entire sky into three parts: Part I region represents the observation field of ZTF, Part II is the observation field of ATLAS excluding the ZTF field and the rest of the sky is indicated by Part III region, which could only be monitored by ASAS-SN. If the randomly generated SN is located in Part I region, we would first determine whether it could be detected by ZTF; if ZTF failed to detect the SN, we try ATLAS and then ASAS-SN. We count this SN as nondetection if it is invisible in all these three surveys. Similarly, if the SN is located in Part II region, we would evaluate whether it could be detected by ALTAS and then by ASAS-SN. For SNe in Part III region, ASAS-SN is the only survey that we will consider. The detection efficiencies of the three surveys are provided in Appendix B. For SNe with different apparent peak magnitudes and light duration, the surveys would detect them at different probabilities. Within the observation window, when an SN was covered by one of the surveys and it was brighter than the detection limit, we could estimate the possibility of this SN being detected by that survey according to the apparent SN magnitude during observation and the detection efficiency given in Appendix B. If there are multiple observations in the considered time window, we would combine the possibilities of each single observation to give an overall evaluation that the SN can be detected (that is, 1-P, where P represents nondetection probability for all observations). By adding up all the possibilities, we could find the fraction of 100,000 SNe in the sample that was detected and the control count according to this fraction. The process was repeated 1,000 times to give the average value and the standard deviation.

The values for $\mathrm{N_{obs}}$ and $\mathrm{N_{ctrl}}$ are given in Table~\ref{rate}, along with the volumetric rates and the estimated statistical and systematic uncertainties.
	
	\begin{table}
		\centering
		\caption{Observed counts, control counts and volumetric rates}\label{rate}
		\begin{threeparttable}
			
			\begin{tabular}{cccc}
				\toprule
				Type & $\mathrm{N_{obs}}$\tablefootmark{a} & $\mathrm{N_{ctrl}}$ & $\mathrm{SNR}$\tablefootmark{b} \\
				\midrule
				Ia & $69.0\pm8.4^{+4.5}_{-3.4}$ & $212.2^{+3.4}_{-4.2}$ & $0.325\pm0.040^{+0.016}_{-0.010}$	\\
				Ibc & $33.5\pm5.8^{+10.0}_{-3.2}$ & $209.8^{+4.5}_{-1.9}$ & $0.160\pm0.028^{+0.044}_{-0.014}$	\\
				II & $108.5\pm10.4^{+37.5}_{-1.3}$ & $205.6^{+7.8}_{-2.5}$ & $0.528\pm0.051^{+0.162}_{-0.013}$	\\
				\bottomrule
			\end{tabular}
			
		\end{threeparttable}
        
         \tablefoot{
         \tablefoottext{a}{For the uncertainties, the first term represents statistical while the second represents systematic uncertainties.}
         \tablefoottext{b}{The volumetric SN rates are in units of $10^{-4} \mathrm{yr^{-1} Mpc^{-3} h^3_{70}}$.}
         }
         
	\end{table}
	
\subsection{Uncertainties}\label{uncertainty}

The uncertainty in the observed count includes the statistical error, which is the Poisson noise for the sample, and the systematic error mainly from three sources:
	(1) "edged-on" SNe with a distance around 40 Mpc;
	(2) SNe with uncertain redshift;
	(3) missing SNe in galaxy cores.
We determined the contribution of the first two sources in the same way as discussed in Paper I. For bright SNe in bright galaxy cores, the presence of the galaxy core will affect the quality of the observed spectra of the SNe. \citet{Desai2024} argued that for surveys such as ASAS-SN, the signal-to-noise ratio of observation of the SNe in the local universe is so high that neglecting the host does not matter for the detection probability. For faint SNe, the noise is dominated by the sky rather than the host, so the presence of a bright host core has little effect on the detection probability. Also, \citet{Holoien2019} reported that quite a few SNe have been discovered in central regions of galaxies by ASAS-SN, that is, at distances even less than 0.02 kpc from the galactic nuclei (see their Fig.2). Such an ability of detecting nuclei SNe or transients holds for ZTF and ATLAS, which could be even better because of having deeper detection limits and better pixel resolutions compared to ASAS-SN. In our SN sample, a large number of SNe located within 1 kpc from the center of their hosts (26 out of a total of 211 SNe) are discovered by ZTF, ATLAS, and ASAS-SN. As the majority of our SN sample was discovered by ASAS-SN, ATLAS and ZTF (including some amateurs), the missing detection of SNe near galactic cores should be less significant for our sample. However, this may not be the case for some amateur surveys, but in fact, all the nearby SNe reported by amateur surveys in our SN sample can be finally captured by the above professional surveys within several days after their reports. 

For the control count we considered the following four sources of systematic error:
(1) The standard deviation of the control count given by the Monte Carlo simulation;
(2) the assumed distribution of host-galaxy dust extinctions;
(3) dust extinction/obscuration that causes a non-negligible fraction of core-collapse SNe to be missed by optical surveys;
(4) uncertainty involved with the assumed models and distributions, including the assumed subtype fractions, distributions of peak absolute magnitudes, light curve templates, differences in magnitudes of different photometric bands, cadence and limit magnitudes of surveys.
The host-galaxy extinction value we generated in the simulation might be underestimated in the cases where the SNe suffer from severe extinctions, then the generated apparent magnitudes could be larger than they should be and cause an overestimation of $\mathrm{N_{ctrl}}$. 
We redid the Monte Carlo simulations by adjusting the host-galaxy extinction distributions according to the extinction data set given by \citet{Holwerda2015} (i.e., increasing high extinction values in Fig. 8 of \citet{Holwerda2015} that the exponential formula we adopted failed to fit). The resultant differences in the derived control count is then set as the uncertainty. Similarly, we could estimate the uncertainty caused by the fourth source according to the uncertainties of the assumed models and distributions. As the effect of the third source was discussed in Paper I, so we directly adopt the value of 57.16 as the number of missed CCSNe and allocate this number to SNe Ibc and II according to their relative proportions to estimate the uncertainty caused by this effect.
	
\subsection{Results}
The final values of the local volumetric rate for SNe Ia, Ibc and II are calculated as 

$$~\mathrm{SNR_{Ia} = 0.325\pm0.040^{+0.016}_{-0.010} \times 10^{-4} yr^{-1} Mpc^{-3} h^3_{70}},$$
	$$\mathrm{SNR_{Ibc} = 0.160\pm0.028^{+0.044}_{-0.014} \times 10^{-4} yr^{-1} Mpc^{-3} h^3_{70}},$$
	$$\mathrm{SNR_{II} = 0.528\pm0.051^{+0.162}_{-0.013} \times 10^{-4} yr^{-1} Mpc^{-3} h^3_{70}},$$respectively. And we combine the last two rates to get the local volumetric rate for CCSNe:
	$$\mathrm{SNR_{CC} = 0.688\pm0.078^{+0.206}_{-0.027} \times 10^{-4} yr^{-1} Mpc^{-3} h^3_{70}}.$$

\section{Supernova rate as a function of galaxy Hubble type} \label{SNuM} 

\subsection{Method}

There are several SN subsamples with different associated galaxy samples in the study of \citet{Li2011b}, including the "full" sample (N = 929), the "full-optimal" sample (N = 726),  the "season" sample (N = 656), the "season-optimal" sample (N = 583), and the "season-optimal" sample (N = 499), respectively. 
\citet{Li2011b} used these samples to calculate the SNuM for SNe Ia, Ibc, and II in a fiducial galaxy of different Hubble types, and the results from different SN subsamples are consistent with each other within 1$\sigma$. Their final rates were calculated using the 726 SNe in the "full-optimal" sample, which provides a good balance between improving statistics of small numbers and avoiding systematic biases.

Our sample size is smaller than that of \citet{Li2011b}, so the control-time method adopted by \citet{Li2011b} is not feasible for us. In Paper I, we used the SNuM and rate-size relation given by \citet{Li2011b} to calculate the expected number of SN explosions in all galaxies of the GLADE+ sample. The rate-size relation is given by

    \begin{equation}
		\mathrm{SNuM(M) = SNuM(M_{0}) (\frac{M}{M_{0}})^{RSS_{M}}}, \label{ratesize}
    \end{equation}

where $\mathrm{M_{0} = 4 \times 10^{10} M_{\odot}}$, is the stellar mass of the fiducial galaxy, the values of $\mathrm{SNuM(M_0)}$\footnote{It represents the SN rate in the fiducial galaxy} and $\mathrm{RSS_M}$ (the rate-size slope, which is the power law index between the rate and the mass) are given in Table 4 of \citet{Li2011b}. Given the stellar mass of any galaxy, we can calculate the corresponding SNuM through Eq.~(\ref{ratesize}), then the expected number of SNe to explode in the galaxy for a given duration of time. Since we have constructed a full galaxy sample in the local universe (i.e., the GLADE+ sample), we could reverse the approach to estimate the SNuM for galaxies of each Hubble type. 

For a given Hubble type, the SN rate for a fiducial galaxy of this type is given by 

    \begin{equation}
		\mathrm{SNuM(M_0) = \frac{N M^{RSS_M}_0}{T \sum M^{RSS_M+1}_i}}, \label{SNuM0}
    \end{equation}

where N is the number of SNe discovered in the given Hubble type of galaxies, $\mathrm{RSS_M}$ is given in the table. 4 of \citet{Li2011b}, T = 8 yr is the duration of the survey and $\mathrm{\sum M^{RSS_M+1}_i}$ is the summation of stellar mass to the power of $(\mathrm{RSS_M} + 1)$ over every galaxy of the given Hubble type in our GLADE+ sample.

\subsection{Uncertainties}

We considered the following sources of uncertainties:
(1) The uncertainty of the number of SNe, the total value of which has already been estimated in Sect.~\ref{Rate}, we just split the total value into the corresponding host galaxy Hubble types;
(2) Uncertainty caused by errors in the RSSs, which has been given in Table 4 of \citet{Li2011b};
(3) The uncertainty of galaxy stellar mass, which has been given in the GLADE+ sample.
The first source is the statistical uncertainty and the rest are systematic uncertainties, we add them and present the total uncertainties in Table~\ref{SNuM_0}.

\subsection{Results}

The rates for galaxies with fiducial size are reported in Table~\ref{SNuM_0} for different Hubble types. In our sample, no CCSNe are found in elliptical galaxies and no SNe Ibc are found in Irr galaxies, so we turn to giving their upper limits. To calculate the rate for a specific galaxy, one simply needs to apply the stellar mass of the galaxy to Eq.~(\ref{ratesize}).

    \begin{table}
		\centering
		\caption{SN rates in fiducial galaxies of different Hubble types.}\label{SNuM_0}
		\begin{threeparttable}
			
			\begin{tabular}{cccc}
				\toprule
				Hubble type & SN type & $\mathrm{SNuM_0}$\tablefootmark{a} & N\tablefootmark{b} \\
				\midrule
				E & Ia & $0.130^{+0.070}_{-0.032}$ & 12	\\
				S0 & Ia & $0.198^{+0.060}_{-0.054}$ & 17 \\
				Sa & Ia & $0.090^{+0.072}_{-0.043}$ & 5	\\
                Sb & Ia & $0.087^{+0.043}_{-0.020}$ & 4	\\
                Sbc & Ia & $0.181^{+0.057}_{-0.025}$ & 5	\\
                Sc & Ia & $0.492^{+0.167}_{-0.120}$ & 18	\\
                Scd & Ia & $0.122^{+0.087}_{-0.055}$ & 4	\\
                Irr & Ia & $0.030^{+0.031}_{-0.035}$ & 1	\\
                 &  &  & 	\\
                E & Ibc & $< 0.010^{+0.006}_{-0.006}$ & 0	\\
                S0 & Ibc & $0.053^{+0.028}_{-0.026}$ & 5	\\
                Sa & Ibc & $0.043^{+0.029}_{-0.039}$ & 2	\\
                Sb & Ibc & $0.220^{+0.087}_{-0.085}$ & 10	\\
                Sbc & Ibc & $0.165^{+0.056}_{-0.095}$ & 6	\\
                Sc & Ibc & $0.122^{+0.088}_{-0.043}$ & 9	\\
                Scd & Ibc & $0.018^{+0.010}_{-0.014}$ & 1	\\
                Irr & Ibc & $< 0.030^{+0.020}_{-0.024}$ & 0	\\
                 &  &  & 	\\
                E & II & $< 0.010^{+0.005}_{-0.004}$ & 0	\\
                S0 & II & $0.043^{+0.018}_{-0.024}$ & 4	\\
                Sa & II & $0.064^{+0.057}_{-0.033}$ & 3	\\
                Sb & II & $0.461^{+0.115}_{-0.150}$ & 21	\\
                Sbc & II & $0.358^{+0.173}_{-0.133}$ & 13	\\
                Sc & II & $0.476^{+0.187}_{-0.122}$ & 35	\\
                Scd & II & $0.265^{+0.125}_{-0.134}$ & 15	\\
                Irr & II & $0.259^{+0.176}_{-0.165}$ & 10	\\
				\bottomrule
			\end{tabular}
			
		\end{threeparttable}

         \tablefoot{
         \tablefoottext{a}{The SN rates are in units of $\mathrm{SN (100 yr)^{-1} (10^{10} M_{\odot})^{-1}}$.}
         \tablefoottext{b}{The number of SNe used in rate calculation.}
         }
         
    \end{table}

\begin{figure}
		\includegraphics[width=\columnwidth]{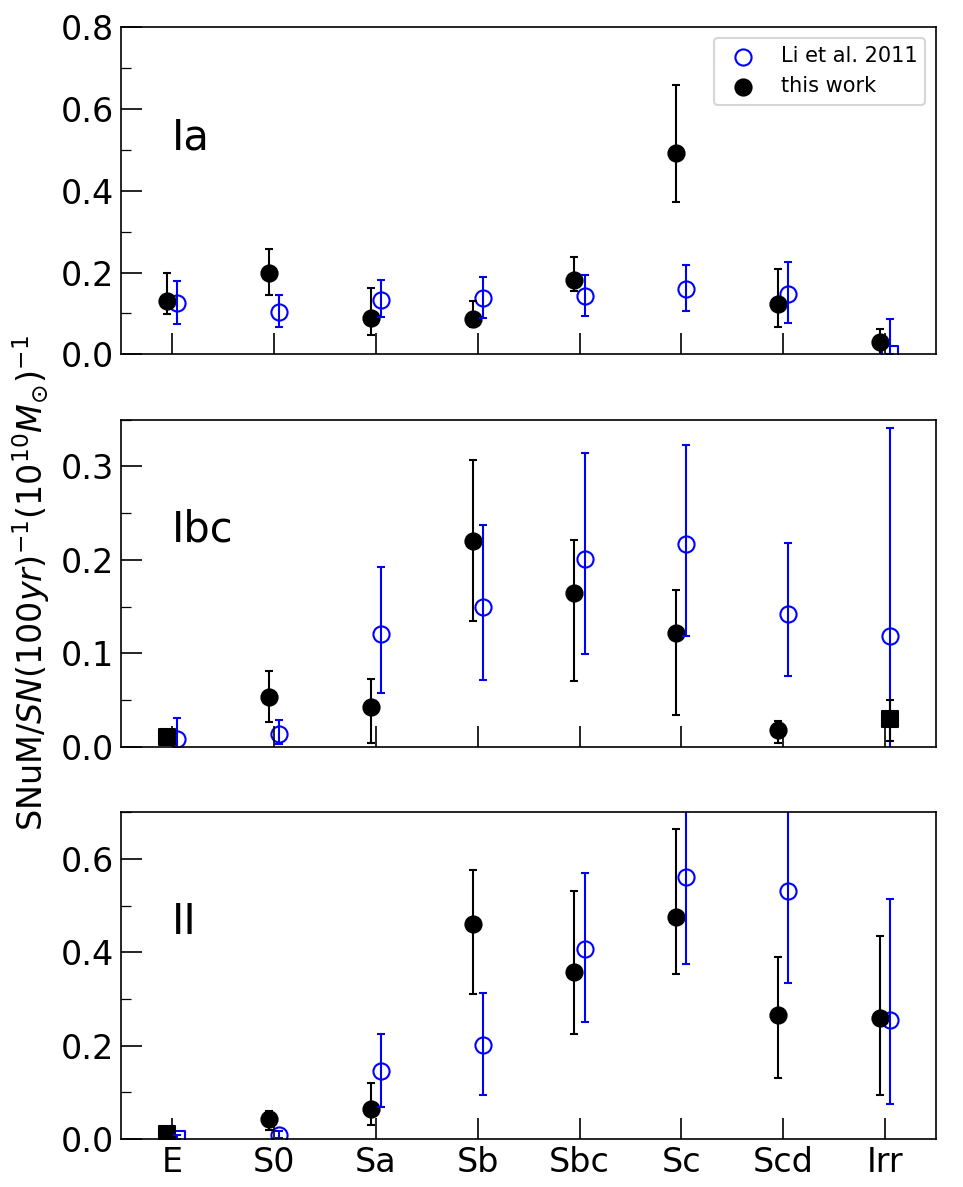}
		\caption{The SN rates (for a galaxy of the fiducial size) for galaxies of different Hubble types (solid circles). The solid squares are upper limits for SNe that are not found in galaxies of the corresponding Hubble type. The blue circles in comparison are the SN rates given by \citet{Li2011b}.}\label{SNuM_fig}
    \end{figure}

All SNuM rates are plotted as solid circles in Fig.~\ref{SNuM_fig}, the upper limits are plotted as solid squares. We also present the rates estimated by \citet{Li2011b} for comparison. 

The SNuM rates of SNe Ia are consistent with \citet{Li2011b}, being constant for different Hubble-type bins except in the Sc galaxies where the rate appears noticeably high. We checked the properties of the Sc galaxies hosting SNe Ia in our sample and found that the uncertainties of their morphological type codes are small (i.e., $<$ 1.0), indicating that their classifications are relatively accurate. Thus, the high SNuM rate of SNe Ia in Sc galaxies could be intrinsic.

In general, the SNuM rates of CCSNe are roughly consistent with the results of \citet{Li2011b}. The CCSNe rates are close to 0 in the E, S0, and Sa galaxies. Then the SN Ibc rates increase to peak in Sb galaxies and gradually decrease to around 0 in Scd and Irr galaxies. This is clearly different from \citet{Li2011b}, where SN Ibc rates peak in Sc galaxies and then drop to a non-zero value in Irr galaxies. SN II rates in galaxies of different Hubble types agree well \citet{Li2011b}, except for the significantly higher rate in Sb galaxies and the lower rate in Scd galaxies. We discussed possible reasons for the small number of CCSNe in Scd and Irr galaxies in our sample in Paper I. It is likely that \citet{Li2011b} overestimated the CCSNe rates in Irr and Scd galaxies due to the preference of massive Scd and Irr galaxies in their observation, and plenty of low-mass Scd and Irr galaxies could be missed in their "full" galaxy sample. The small sample size of our SN Ibc sample might also contribute to the discrepancy. Figs. 8 and 9 in Paper I show that the average stellar mass of Sb galaxies in the SN-host galaxy sample is smaller than that in the GLADE+ sample. Furthermore, the number of CCSNe discovered in Sb galaxies is also larger compared to \citet{Li2011b}. This suggests that more CCSNe tend to explode in less massive galaxies, which naturally results in a higher CCSN rate in Sb galaxies. Notice that for the SNuM calculation in terms of the "control-time method", a large sample size is needed.

\subsection{The Galactic SN rate}

We tried to determine the expected SN rate in the Milky Way (hereafter, the Galactic SN rate) from the SN rates we derived from the 40-Mpc sample. We assumed that the Hubble type of Milky Way is Sbc \citep{van1994}. For the stellar mass, we adopted the value of $4.81 \pm 0.13 \times 10^{10} \mathrm{M_{\odot}}$ \citep{Lian2024}. In comparison, according to \citet{Li2011b}, we can assume that the Milky Way has a size similar to that of the Andromeda galaxy (M31) or the average size of the Sbc galaxies in the “optimal” LOSS galaxy sample of \citet{Leaman2011}. Then we can give the Galactic SN rate in units of SNe per century using the rate-size relation of Eq.~(\ref{ratesize}). The relevant results are given in Table~\ref{Galactic}.

The Galactic SN rate we estimate from the stellar mass of the Milky Way given by \citet{Lian2024} is $3.08 \pm 1.29$ SNe per century, just in between the estimations from the stellar masses suggested by \citet{Li2011}, and consistent with the value of $2.84 \pm 0.60$ SNe per century obtained by \citet{Li2011b}. Our result is also in good agreement with the published results of 1.4$-$5.8 SNe per century based on different methods.
 
    \begin{table*}
		\centering
		\caption{Galactic SN rate\tablefootmark{a}.}\label{Galactic}
		\begin{threeparttable}
		  \resizebox{\textwidth}{!}{
			\begin{tabular}{ccccccc}
				\toprule
				Mass\tablefootmark{b} & Ia & Ibc & II & CCSNe & Total SNe & Comments \\
				\midrule
				4.81 & $0.80 \pm 0.23$ & $0.72 \pm 0.35$ & $1.56 \pm 0.71$ & $2.27 \pm 1.06$ & $3.08 \pm 1.29$ & (1)	\\
				2.3 & $0.53 \pm 0.07$ & $0.51 \pm 0.29$ & $1.12 \pm 0.58$ & $1.63 \pm 0.87$ & $2.16 \pm 0.94$ & M31 \\
				5.2 & $0.84 \pm 0.23$ & $0.74 \pm 0.35$ & $1.61 \pm 0.71$ & $2.35 \pm 1.05$ & $3.19 \pm 1.29$ & Average Sbc galaxy	\\
				\bottomrule
			\end{tabular}
	    }	
			
		\end{threeparttable}

        \tablefoot{
         \tablefoottext{a}{In unit of SNe per century.}
         \tablefoottext{b}{Stellar mass in units of $10^{10} \mathrm{M}_{\odot}$.}
         }
         \tablebib{(1) \citet{Lian2024}.}
    \end{table*}

\section{Analysis} \label{analysis}
	
\subsection{Comparison with historical results}
	
We first compare our SN Ia and CCSN rates with other historical results. Most were estimated with observation data from the targeted surveys, such as \citet{Li2011b}. These surveys tend to monitor brighter, more massive galaxies, thus introducing bias in the observed SN population, as the light curves of SNe and their host galaxy properties are correlated \citep{Sullivan2010}. The final volumetric rates would also suffer from such a bias. For comparison, we also include recent results of untargeted rolling searches, for instance, PTF \citep{Frohmaier2019} and ASAS-SN \citep{Desai2024}. These results do not suffer from observational bias as the targeted surveys, thus could achieve better precision.
	
\subsubsection{Type Ia Supernovae}

\begin{figure}
		\includegraphics[width=\columnwidth]{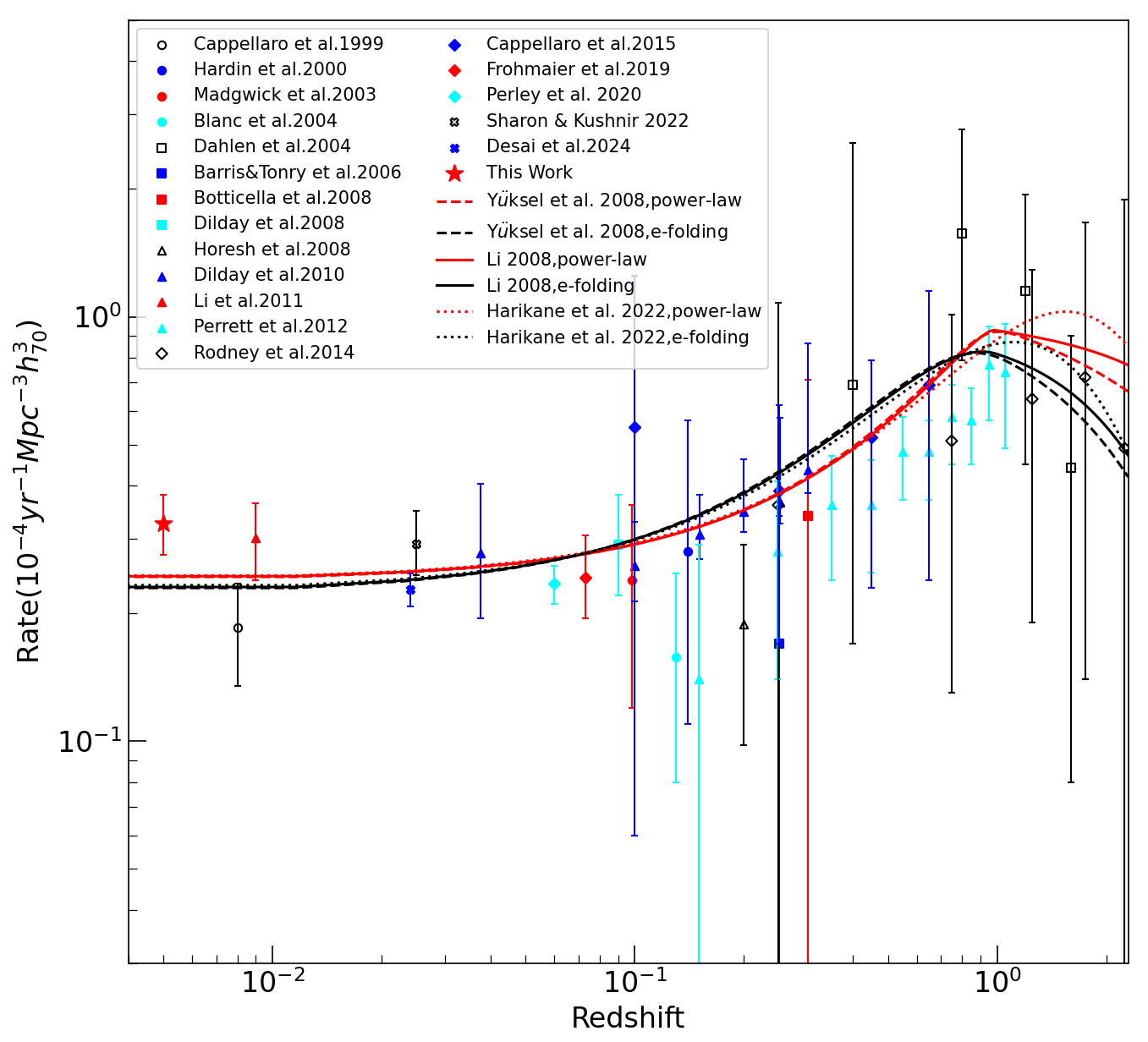}
		\caption{Volumetric SN Ia rate compared to historical results at different redshifts. The red and black lines show the SN Ia rates derived from power-law and e-folding form of DTD, respectively. The dashed, solid and dotted lines show the SN Ia rates derived from SFH given by \citet{Yuksel2008}, \citet{Li2008} and \citet{Harikane2022}, respectively.}\label{Ia_rate}
    \end{figure}

Comparison of our SN Ia rate to other survey results is shown in Fig.~\ref{Ia_rate}. We include published rates of \citet{Cappellaro1999}, \citet{Hardin2000}, \citet{Madgwick2003}, \citet{Blanc2004}, \citet{Dahlen2004}, \citet{Barris2006}, \citet{Botticella2008}, \citet{Dilday2008}, \citet{Horesh2008}, \citet{Dilday2010}, \citet{Li2011b}, \citet{Perrett2012}, \citet{Rodney2014}, \citet{Cappellaro2015}, \citet{Frohmaier2019}, \citet{Perley2020}, \citet{Sharon2022}, \citet{Desai2024} and adjust them to the assumed cosmological model. A summary of the local SN Ia rate measurement is shown in Table~\ref{ratesum}, all rate measurements used in this work are provided in Table~\ref{Iarateall}.
	
Our rate is plotted as a red star with a value slightly larger than the results given by \citet{Li2011b} at a confidence level of about 1$\sigma$. However, we achieve better precision (smaller uncertainty) compared to most local rate estimations, similar to the recent values of \citet{Frohmaier2019}. The SN Ia rate experiences a rapid decline from redshift z = 0 to z $\sim$ 0.1, followed by a gradual increase until z $\sim$ 1, and then decreases. We notice that there exists an unusual rate decrease from z = 0 to z $\sim$ 0.1, we will discuss the possible explanations in Sects.~\ref{DTD} and ~\ref{metal}

    \begin{table*}
		\centering
		\caption{Local volumetric SN rates.}\label{ratesum}
		\begin{threeparttable}
		
			\begin{tabular}{cccccccc}
				\toprule
                \multicolumn{4}{c}{SN Ia} & \multicolumn{4}{c}{CCSN} \\
				z & N$_{\mathrm{SN}}$ & Rate$/10^{-4} \mathrm{yr^{-1} Mpc^{-3} h^3_{70}}$ & Reference & z & N$_{\mathrm{SN}}$ & Rate$/10^{-4} \mathrm{yr^{-1} Mpc^{-3} h^3_{70}}$ & Reference \\
				\midrule
				$\sim$ 0 & 69.6 & $0.185 \pm 0.05$ & (2) & $\sim 0$ & 440 & $0.705 \pm 0.211$ & (10) \\
				0.098 & 19 & $0.24 \pm 0.12$ & (11) & 0.003 & 14 & $1.6 \pm 0.4$ & (1) \\
				0.09 & 17 & $0.29^{+0.09}_{-0.07}$ & (5) & 0.072 & 89 & $1.06 \pm 0.26$ & (14) 	\\
                0.025-0.050 & 4 & $0.278^{+0.127}_{-0.083}$ & (6) & 0.075 & 16 & $1.04 \pm 0.37$ & (9) \\
                $\sim 0$ & 274 & $0.301 \pm 0.062$ & (10) & 0.05-0.15 & 50 & $1.13^{+1.11}_{-1.02}$ & (3) 	\\
                0.05-0.15 & 3 & $0.55^{+0.70}_{-0.49}$ & (3) & 0.028 & 86 & $0.910^{+0.156}_{-0.127}$ & (8)	\\
                0.073 & 90 & $0.243^{+0.062}_{-0.048}$ & (7) & 0-0.01 & 142 &  $0.688^{+0.284}_{-0.105}$ & this work	\\
                < 0.1 & 875 & $0.235 \pm 0.024$ & (12) &  &  &  &	\\
                0.01-0.04 & 298 & $0.291^{+0.058}_{-0.045}$ & (13) &  &  &  &	\\
                0.024 & 404 & $0.228 \pm 0.020$ & (4) &  &  &  &	\\
                0-0.01 & 69 & $0.325^{+0.056}_{-0.050}$ & this work &  &  &  &	\\
				\bottomrule
			\end{tabular}
            	
		\end{threeparttable}

        \tablebib{(1) \citet{Botticella2012}; (2) \citet{Cappellaro1999}; (3) \citet{Cappellaro2015}; (4) \citet{Desai2024}; (5) \citet{Dilday2008}; (6) \citet{Dilday2010}; (7) \citet{Frohmaier2019}; (8) \citet{Frohmaier2021}; (9) \citet{Graur2015}; (10) \citet{Li2011b}; (11) \citet{Madgwick2003}; (12) \citet{Perley2020}; (13) \citet{Sharon2022}; (14) \citet{Taylor2014}.}
        
    \end{table*}
 
\subsubsection{Core-collapse supernovae}
	
\begin{figure}
		\includegraphics[width=\columnwidth]{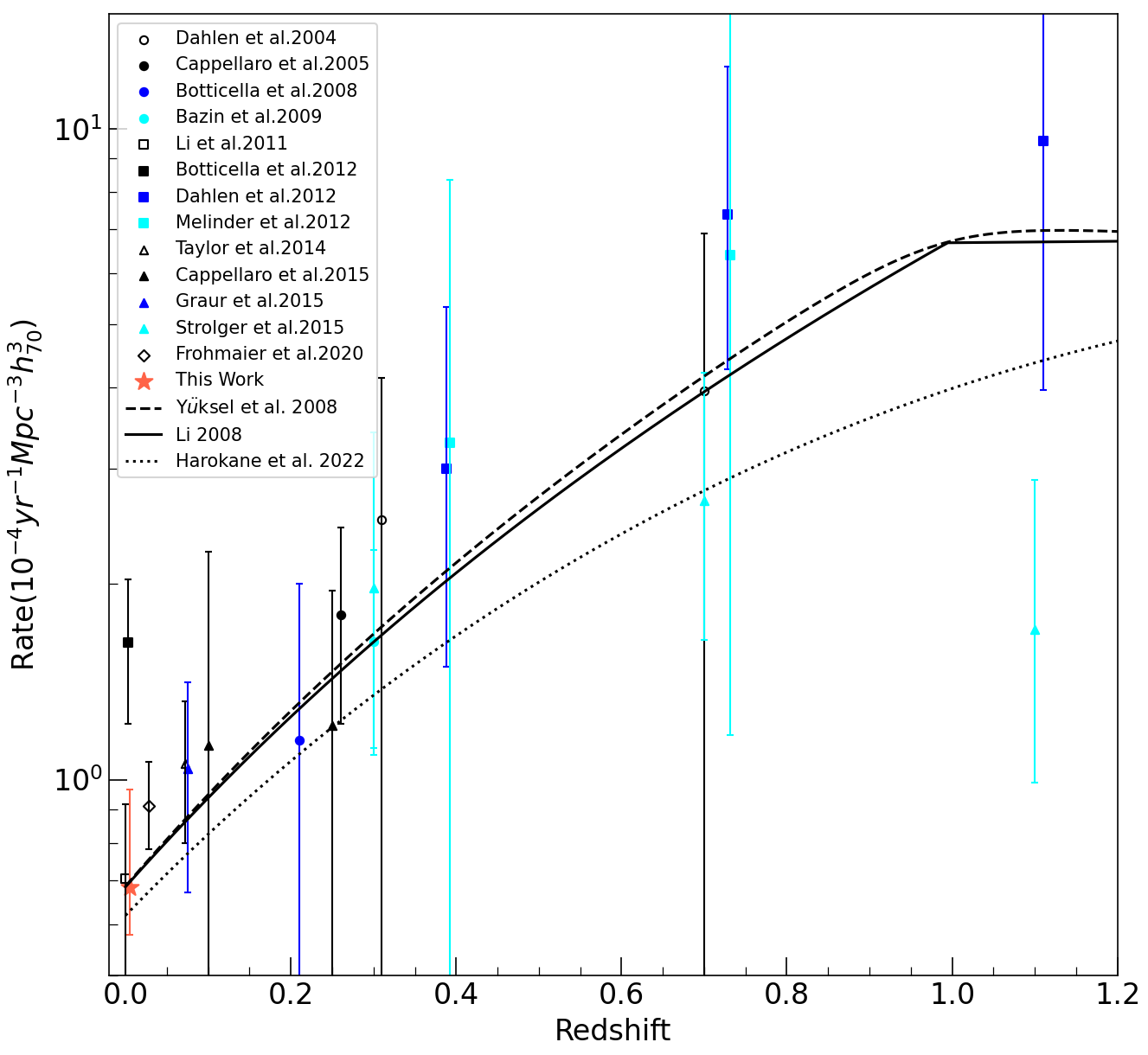}
		\caption{The same as Figure \ref{Ia_rate}, but for CCSNe. 
        The dashed, solid and dotted black lines show the CCSN rates derived from SFH given by  \citet{Yuksel2008}, \citet{Li2008} and \citet{Harikane2022}, respectively.}\label{CC_rate}
    \end{figure}

Our CCSN rate is compared to the historical results of \citet{Dahlen2004}, \citet{Cappellaro2005}, \citet{Botticella2008}, \citet{Bazin2009}, \citet{Li2011b}, \citet{Botticella2012}, \citet{Dahlen2012}, \citet{Melinder2012}, \citet{Taylor2014}, \citet{Cappellaro2015}, \citet{Graur2015}, \citet{Strolger2015}, and \citet{Frohmaier2021}. The published values, scaled to the adopted value of the Hubble constant, are shown in Fig.~\ref{CC_rate}. A summary of local CCSN rate measurements is shown in Table~\ref{ratesum}, and all rate measurements used in this work are provided in Table~\ref{IIrateall}.
	
Our rate is well consistent with that given by \citet{Li2011b} but with slightly higher precision. 
	
\subsection{The delay-time distribution} \label{DTD}
    
In this section, we introduce the application of our improved nearby SN Ia rate measurement. The evolution of SN Ia rate should follow the star formation history (SFH), but because of the long evolution timescale of their progenitor system, the SN Ia DTD should be taken into account. The SN Ia rate can be modeled as the convolution of DTD and SFH, that is, 
	\begin{equation}
		\mathrm{SNR_{Ia}}(t) = \mu \int_{t}^{t_F} \mathrm{SFH}(t')\Psi(t'-t)\,dt', \label{Ia}
	\end{equation}
where $\mathrm{t'-t}$ is the delay time, t is the look-back time corresponding to the redshift at which we evaluate the SN Ia rate, t$_\mathrm{F}$ is the look-back time corresponding to the redshift, z$_\mathrm{F}$, where the first stars formed. We set z$_\mathrm{F} = 10$. $\mu$ is the scale factor and $\Psi(t)$ is the DTD. 
 
First, we used a power-law DTD, $\Psi(t) = t^{-\beta}$, and an e-folding DTD, $\Psi(t) = exp(-t/\tau)$, to fit our data. $\tau$ is the characteristic delay time for the e-folding DTD. For the SFH, we adopted the functional form of \citet{Yuksel2008} and \citet{Harikane2022}, and the piece-wise form of \citet{Li2008} given in Eqs.~(\ref{SFH1}), ~(\ref{SFH2}), and ~(\ref{SFH3}), respectively. 
	\begin{equation}
		\mathrm{SFH}(z) = \mathrm{SFH}_0 [(1 + z)^{a\eta} + (\frac{1 + z}{\mathrm{B}})^{b\eta} + (\frac{1 + z}{\mathrm{C}})^{c\eta}]^{\frac{1}{\eta}}, \label{SFH1}
	\end{equation}
	with a = 3.4, b $= -0.3$, c $= -3.5$, $\mathrm{SFH_0 = 0.02 M_\odot yr^{-1} Mpc}^{-3}$, B $\simeq 5000$, C $\simeq 9$, and $\eta \simeq -10$.
    \begin{equation}
		\mathrm{SFH}(z) = \frac{h \mathrm{M_\odot yr^{-1} Mpc}^{-3}}{61.7 \times (1+z)^{-3.13} + 10^{0.22(1+z)} + 2.4 \times 10^{0.50(1+z)-3.0}}, \label{SFH2}
	\end{equation}
    \begin{equation}
		log\mathrm{SFH}(z) = \mathrm{a + b} log(1+z), \label{SFH3}
	\end{equation} 
	$$(a,b) = \left\{ \begin{array}{lcl}
		(-1.70, 3.30) & \mbox{for} & z < 0.993 \\
		(-0.727, 0.0549) & \mbox{for} & 0.993 < z < 3.80. \\
		(2.35, -4.46) & \mbox{for} & z > 3.80
	\end{array}\right.$$
The resulting best-fit parameters are listed in Table~\ref{para1} and the SNRs derived from Eq.~(\ref{Ia}) using different SFH models are shown in Fig.~\ref{Ia_rate}.
	
	\begin{table*}
		\centering
		\caption{Best-fit parameters for the power-law and e-folding DTD convolved with different SFH models.}\label{para1}
		\begin{threeparttable}
			
			\begin{tabular}{ccccccc}
				\toprule
                & \multicolumn{3}{c}{power-law} & \multicolumn{3}{c}{e-folding} \\
				SFH model & $\mu/\mathrm{yr M_\odot^{-1}}$ & $\beta$ & $\chi^2$ & $\mu/\mathrm{M_\odot^{-1}}$ & $\tau/$Gyr & $\chi^2$\\
				\midrule
				\citet{Yuksel2008} & $0.75 \pm 0.09$ & $1.08 \pm 0.05$ & 1.26 & $28.57 \pm 3.85$ & $2.39 \pm 0.22$ & 1.42	\\
				\citet{Li2008} & $0.75 \pm 0.09$ & $1.08 \pm 0.05$ & 1.40 & $27.93 \pm 3.75$ & $2.46 \pm 0.23$ & 1.36	\\
                \citet{Harikane2022} & $0.89 \pm 0.17$ & $1.17 \pm 0.07$ & 1.88 & $50.96 \pm 7.80$ & $2.12 \pm 0.23$ & 1.30	\\
				\bottomrule
			\end{tabular}
			
		\end{threeparttable}
	\end{table*}
	
All models presented in Fig.~\ref{Ia_rate} fit well with the overall rate measurements. All power-law models give $\beta \sim 1$, consistent with historical fittings (i.e., $\beta = 1.13 \pm 0.05$ from \citet{Wiseman2021}). This power-law DTD is consistent with a progenitor scenario of DD\citep{Maoz2012,Graur2013}. All e-folding models favor a characteristic delay timescale of around 2 Gyr. \citet{Strolger2020} also derived a family of delay time distribution solutions from the volumetric evolution SN Ia rate and suggested an exponential distribution similar to the $\beta \sim 1$ power-law distribution, consistent with our results.
However, none of these models are capable of matching the short decline in SN Ia rate observed from the universe at z = 0 to that at $z \simeq 0.1$ universe. 
The recent rates measured at $<z> = 0.024$ and 0.073 from the ASAS-SN and PTF projects have relatively high precision, so this short decline could be an intrinsic trend. Inspired by the double peak structure identified in Paper I, we further examined another DTD model, the two-component model.

\begin{figure}
        \includegraphics[width=\columnwidth]{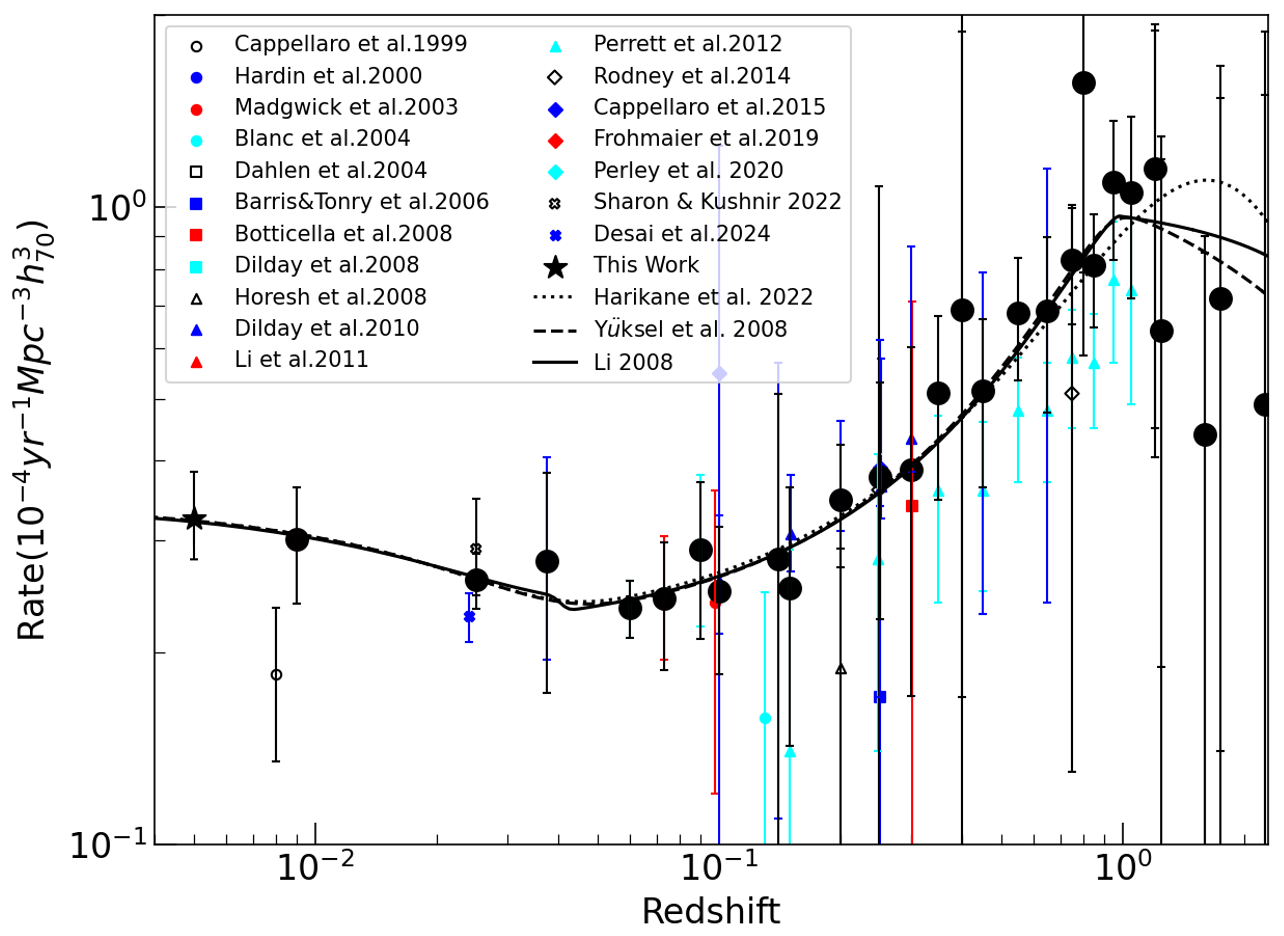}
		\caption{The SN rates derived using different SFH models and best-fit parameters of the two-component model. The dashed, solid and dotted lines correspond to the SFH given by \citet{Yuksel2008}, \citet{Li2008} and \citet{Harikane2022}, respectively. The black dots show the binned rate measurements used in our DTD fitting procedure.}\label{DG}
        \end{figure}

The most popular two-component model is the "A+B" model which is composed of a "prompt" component that tracks the instantaneous star formation and a "delayed" component that is proportional to $\mathrm{M_{stellar}}$\citep{Mannucci2005}:
        \begin{equation}
            \mathrm{SNR_{Ia}(t) = A \times M_{stellar}(z) + B} \times \mathrm{SFH}(z) \label{A+B}
        \end{equation}
The A and B coefficients scale the "delayed" and "prompt" components, respectively. However, the "delayed" component of this A+B model needs to be converted to a DTD by using the relation between $\mathrm{M_{stellar}}$ and time. The "prompt" component consists of SNe Ia that explode immediately after the formation of their progenitor systems with no delay time at all, since it is proportional to the SFH. But this zero delay time is unrealistic for SNe Ia. Since the DTD of the power law function could fit the evolution of the SN Ia rate from z $\sim$ 0.1 to larger redshifts, we turned to another functional form of DTD, which consists of a power-law component plus a Gaussian component. The power-law component contributes to the SN Ia rate at large redshifts, while the Gaussian component is the correction for the decline in SN Ia rate in the local universe:
        \begin{equation}
            \Psi(t) = \frac{a}{\sqrt{2\pi}\sigma} exp(-\frac{(t-\tau)^2}{2\sigma^2}) + b (t)^{-\beta},\label{doubleGaussian}
        \end{equation}
where $\tau$ is the mean delay time of the Gaussian component, a and b represent the normalization parameters. The corresponding best fit parameters are listed in Table~\ref{para2} and the SNRs derived from Eq.~(\ref{Ia}) using different SFH models are shown in Fig.~\ref{DG}.

        \begin{table*}
		\centering
		\caption{Best-fit parameters for the two-component models.}\label{para2}
		\begin{threeparttable}

			\begin{tabular}{ccccccc}
				\toprule
				SFH model & a & $\tau/$Gyr & $\sigma/$Gyr & b & $\beta$ & $\chi^2$\\
				\midrule
				\citet{Yuksel2008} & 0.70 & $ 12.63 \pm 0.38 $ & $ 0.20 \pm 0.38 $ & 0.54 & $1.18 \pm 0.04$ & 1.23 \\ 
				\citet{Li2008} & 0.57 & $ 12.45 \pm 0.15$ & $ 0.02 \pm 0.46$ & 0.54 & $1.18 \pm 0.04$ & 1.41	\\ 
                \citet{Harikane2022} & 0.97 & $ 12.48 \pm 0.53$ & $0.14 \pm 0.71$ & 0.58 & $1.29 \pm 0.08$ & 1.98	\\ 
				\bottomrule
			\end{tabular}

		\end{threeparttable}
	\end{table*}

\begin{figure}
        \includegraphics[width=\columnwidth]{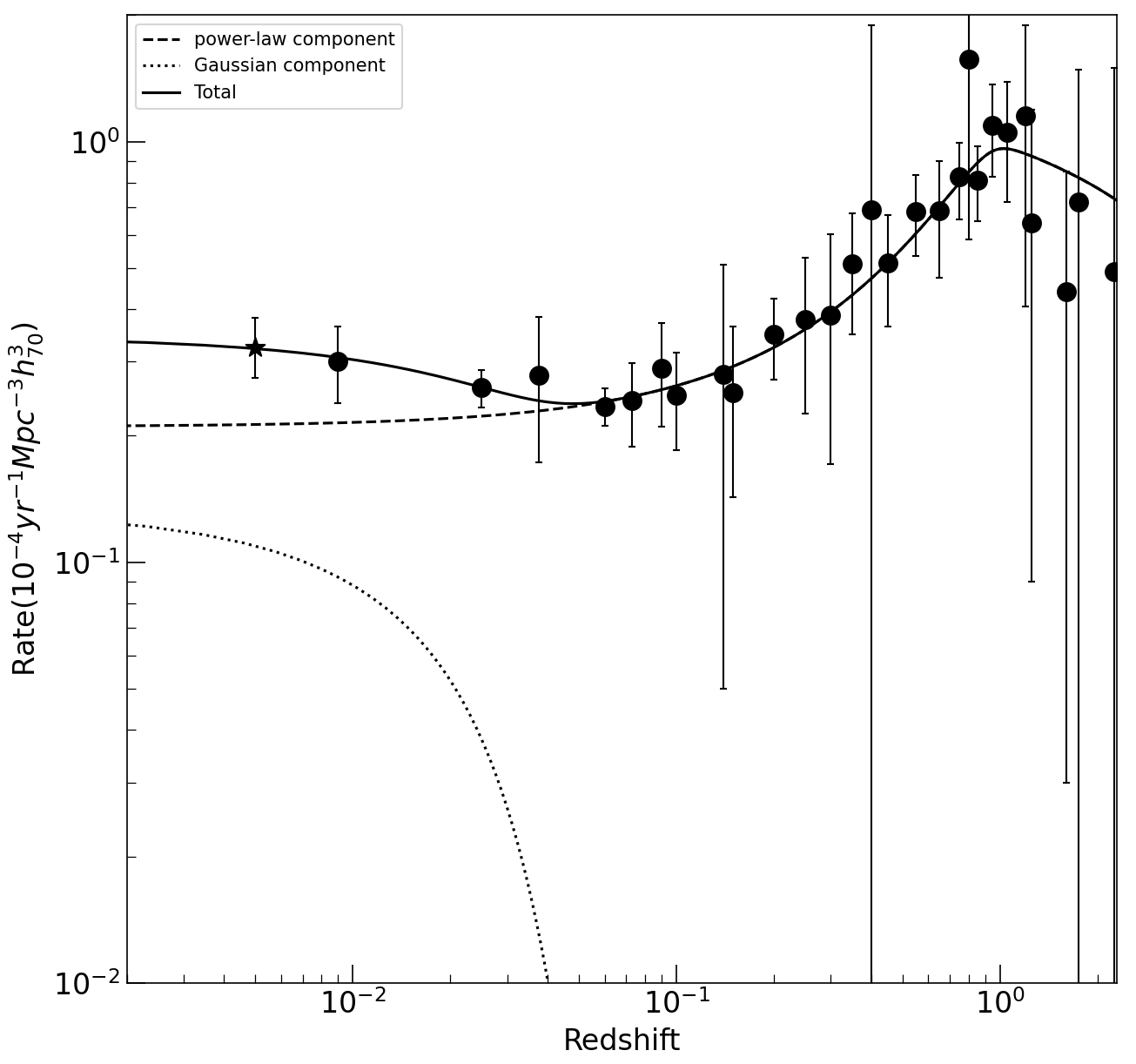}
		\caption{The two-component model of \citet{Yuksel2008} SFH. The solid line represents the best-fit evolution of SN Ia rate while the dashed and dotted lines correspond to the power-law (prompt) and Gaussian (delayed) components, respectively.}\label{Y}
    \end{figure}

All three models provide a proper fit to the evolution of SN Ia rate at z $>$ 0.1, as well as the rate decline in the local universe, whereas a single-component DTD model (i.e., the power law and Gaussian DTD model) failed to give a reasonable fit. 
The power-law components all give $\beta \sim 1$, consistent with the historical results of the single power-law DTD model, which corresponds to the prompt part of the two-component model with delay times close to 0. The delayed component of all three models has a delay time of $\sim$ 12.5 Gyr, we choose the best fit value of \citet{Yuksel2008} model, which gives $\beta = 1.18 \pm 0.04$ and $\tau = 12.63 \pm 0.38$ (corresponding to a redshift of z $\sim$ 6 for the assumed cosmological parameters). This result could suggest a large number of star-forming galaxies at $z \sim 6$. We present details of the two-component model of \citet{Yuksel2008} SFH in Fig.~\ref{Y}. The unusual rate increase from z $\sim$ 0.1 to the local universe can be attributed to the Gaussian component with a long delay time. 

Recent observations of JWST have uncovered an increasing number of galaxies at $z > 6$ \citep{Jaskot2024}, with some having redshifts beyond 12 and out to 14 \citep{Chakraborty2024,Lu2025}. The JWST Advanced Deep Extragalactic Survey (JADES) discovered a sample of 79 SNe in the JADES Deep Field, which contains many high-redshift SNe, with 7 at z $\geq$ 4, 15 at z $\geq$ 3 and 38 at z $\geq$ 2 \citep{DeCoursey2025}. The sample includes a spectroscopically-confirmed SN Ia at z = 2.90 \citep{Pierel2024}, and a SN IIP at z = 3.61, which represent the highest-redshift SNe Ia and SN IIP ever discovered. The production of SNe at high redshifts suggests star formation activities in evolved galaxies at even higher redshifts. Moreover, historical studies on the host galaxies of SNe Ia have shown that their metallicity is systematically higher than that of CCSNe \citep{Shao2014,Galbany2016}. The relatively high metallicity environment suggests that the birth of zero-metallicity first-generation stars could be much earlier, considering the non-negligible delay time of SNe Ia. Thus a delay time as long as 12.63 Gyr could be possible. In summary, the observation results of JWST favor the presence of large delay times in the formation of SNe Ia as suggested by our analysis. We will further investigate the possible origin of the rate decline in the local universe in Sect.~\ref{metal}.

The outcome of the DTD fitting can provide insight into the progenitor systems of SNe Ia. The DTD of the SD channel has a sharp cutoff at 2-3 Gyrs \citep{Han2004}. From the models of \citet{Wang2010}, SNe Ia from the SD models have a wide delay-time distribution, where the WD $+$ He star channel contributes to the SNe Ia with delay times shorter than 100 Myr, and the WD $+$ MS and WD $+$ RG channels to those with age longer than 1 Gyr, but with a very small fraction longer than 3 Gyr (the WD $+$ MS channel also contributes to the SNe Ia with intermediate delay times around 100 Myr-1 Gyr), so SD channel cannot account for our delayed component but still has a potential contribution to the prompt component. For the DD channel, the models have given a delay time ranging from several Gyrs up to around 10 Gyr \citep{Pakmor2013,Crocker2017,Perets2019}. The model simulations of \citet{Ruiter2009,WangB2010} showed that the SNe Ia from the DD channel has delay times of a few Myr $-$ 15 Gyr. The DTD of the DD channel peaks at several hundred Myrs, followed by a decrease in power law $\propto t^{-1}$, which corresponds to the power law component of our model. The tail could reach $\sim$15 Gyr with an event rate of around 2 orders of magnitude lower than the peak value. From their model simulations, we can conclude that the DD channel could account for both the prompt and the delayed components of our two-component DTD model. However, these models all give a power law decline for the DD channel, with a very small event rate for delay times over 10 Gyr, so these models suggest that the long delay time of 12.63 $\pm$ 0.38 Gyr is possible for the DD progenitor systems, but none of them could explain the Gaussian component existing at large delay times of our model. \citet{Briel2022} estimated SN Ia rate using models of Binary Population And Spectral Synthesis \citep[BPASS,][]{Eldridge2017,Stanway2018} and derived the DTD of SNe Ia at different metallicities (see their Fig. 1). At solar metallicity, the delay time clearly exists a second peak at an age larger than 10 Gyr. \citet{Joshi2024} recovered the DTD of SNe Ia in different host galaxy samples and found that, for host galaxies with zero star formation at look-back time less than 10 Gyr, a significant component exists in the DTD bin around 12 Gyr (see their Fig. 3). Both studies indicated the existence of a group of SNe Ia with extremely long delay times. However, further investigation of the stellar evolution models of the DD channel is encouraged to explore this problem.

\subsection{The star formation rates} \label{SFH}
	
For CCSNe, the relationship between their birth rate and SFH is straightforward. Since their progenitors are massive stars with relatively short lifetimes, the SNR of CCSNe should be proportional to the SFH. 
	
Assuming a Salpeter initial mass function \citep[IMF;][]{Salpeter1955}, $\psi(\mathrm{M})$, in the range $0.1<\mathrm{M/M}_\odot<125$ and all stars with masses in the range $8<\mathrm{M/M}_\odot<50$ exploding as SNe, we can get the relation between SFH (in units of $\mathrm{M_\odot yr^{-1} Mpc^{-3}}$) and SNR of CCSNe (in units of $\mathrm{yr^{-1} Mpc^{-3}}$), 
	\begin{equation}
		\mathrm{SNR_{CC}}(z) = k \times \mathrm{SFH}(z), \label{CC}
	\end{equation}
	where $$\mathrm{k = \frac{\int_{8M_\odot}^{50M_\odot}\psi(M)\,dM}{\int_{0.1M_\odot}^{125M_\odot}M\psi(M)\,dM} = 0.0070M_\odot^{-1}}.$$
	
We normalized SFH as parameterized by \citet{Yuksel2008}, \citet{Li2008} and \citet{Harikane2022} using Eq.~(\ref{CC}). Fig.~\ref{CC_rate} shows the resulting SN rate. 
The CCSN rate has the same trend as the SFH given by \citet{Yuksel2008} and \citet{Li2008} from z = 0 to around 1 (a look-back time of $\sim$ 7.7 Gyr), consistent with the conclusion of \citet{Briel2022}. But the result of \citet{Harikane2022} SFH significantly underestimates the CC rates. Our local rate measurement is perfectly consistent with the cosmic SFH from the literature. Further improvements in our understanding of SFH would come from better-constrained rate measurements of high-redshift ($z > 1$) CCSNe. 

Both \citet{Mannucci2007} and \citet{Mattila2012} suggested that a large fraction of CCSNe would be missed by optical surveys even in the nearby universe due to dust obscuration. \citet{Mattila2012} gave the evolution of the missing CCSNe fraction with respect to redshift, from $18.9^{+19.2}_{-9.5}\%$ at z = 0 to $35.9^{+21.0}_{-9.0}\%$ at z = 2.0 (see their Table 10 for the details). Although the surveys like ZTF, ATLAS, and ASAS-SN have recently gained significant improvements in detecting SNe in a systematic way, the effect of dust obscuration might still be non-negligible, even at small distances. \citet{Jencson2018} discovered SPIRITS 16tn, a type II SN discovered with Spitzer/IRAC during the ongoing SPIRITS (SPitzer InfraRed Intensive Transients Survey) survey in the nearby galaxy NGC 3556. The SN is only 8.8 Mpc away from us, but it was completely missed by optical searches due to heavy extinction. The birth rates of CCSNe would then be systematically higher than those estimated from the cosmic star-formation history.

\subsection{Comparison to the PTF SN Ia sample}\label{metal} 

To further investigate the possible origin of the unusual evolution of SN Ia rate in the local universe, we compared the host environments of our SN Ia sample with those of the \citet{Frohmaier2019} sample from the Palomar Transient Factory (PTF) survey (with an average redshift of 0.073). We estimated the host galaxy properties (i.e., host metallicity and stellar mass) of the PTF sample following the same method as in Sect. 4.2 of Paper I.

    \begin{figure}
		\includegraphics[width=\columnwidth]{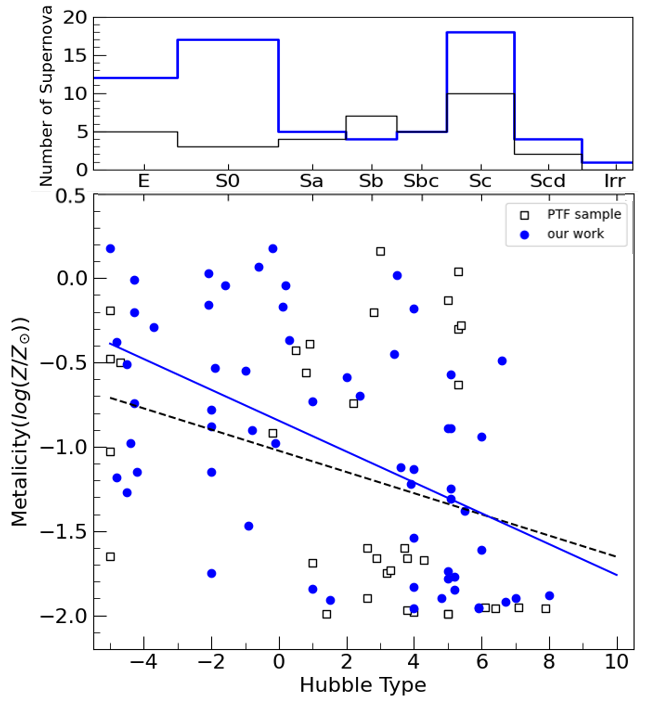} 
		\caption{Upper:Number of SNe in galaxies of different Hubble types for our sample (blue) and the PTF sample (black). Lower:Correlations between host galaxy metallicity and Hubble type for our SN Ia sample (blue dots). We also show the correlation for PTF sample as comparison (black squares). The solid blue line and dashed black line represent the linear fitting of our sample and PTF sample, respectively.}\label{tZ}
    \end{figure}

    \begin{figure}
		\includegraphics[width=\columnwidth]{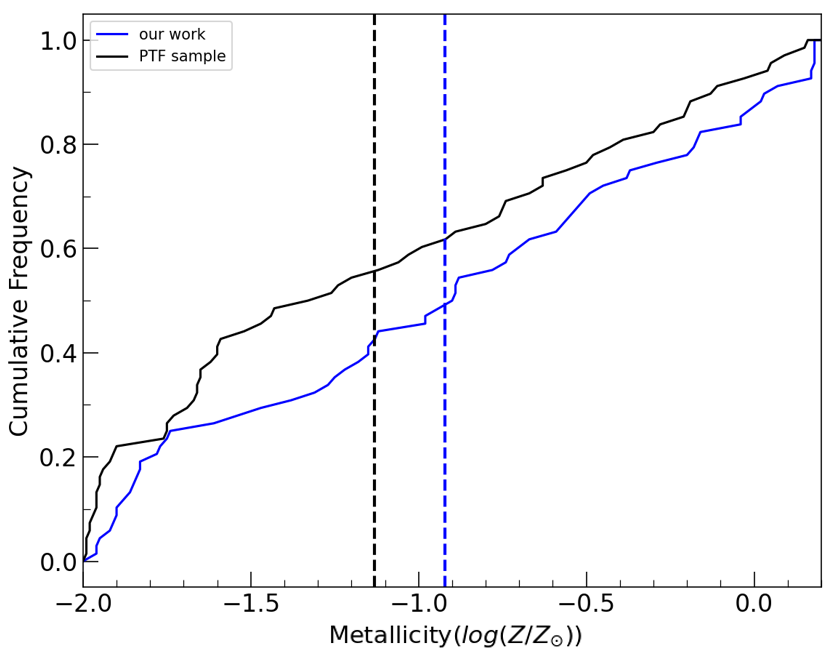}
		\caption{The cumulative fractions for the host metallicity distributions of our SN Ia sample (blue) and the sample of \citet{Frohmaier2019}. The vertical dashed lines show the average values of the host metallicity for the two SNe Ia samples.}\label{metallicity}
    \end{figure}

    \begin{figure}
		\includegraphics[width=\columnwidth]{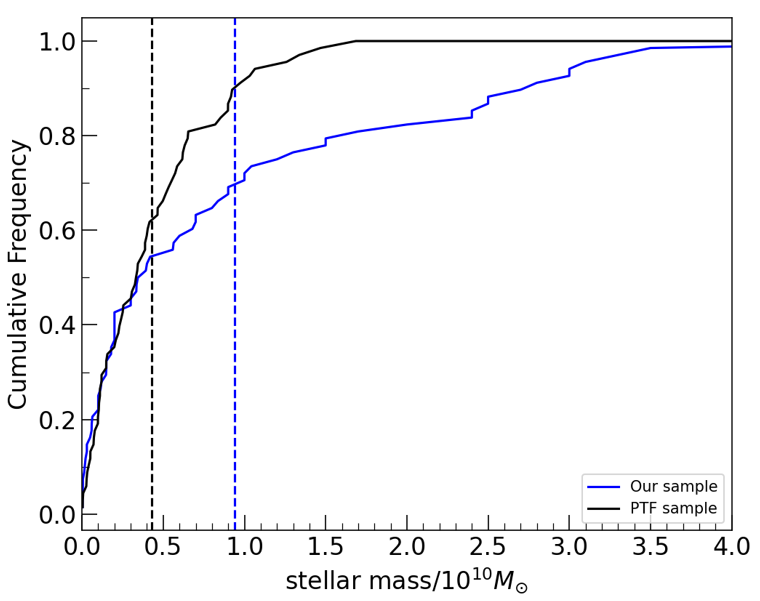}
		\caption{The cumulative fractions for the host stellar mass distributions of our SNe Ia sample (blue) and that of \citet{Frohmaier2019}. The vertical dashed lines show the average values of the host stellar mass for two SNe Ia samples.}\label{mass}
    \end{figure}
    
The correlation between host metallicity and Hubble type is shown in Fig.~\ref{tZ}, we did a linear fit of the two samples and they show clear difference that the fraction of E-S0 galaxies in our sample is significantly higher, resulting in higher metallicity at the elliptical end. The slopes of the two lines are $-0.09 \pm 0.02$ for our sample and $-0.06 \pm 0.03$ for the PTF sample, respectively. Moreover, the PTF SNe Ia sample shows a different Hubble-type distribution with a clear preference to reside in spiral galaxies rather than E-S0 galaxies (seen in the upper panel of Fig.~\ref{tZ}). A K-S test gives the p value as 0.04, suggesting that this difference is statistically significant. The lack of SNe Ia in metal-rich E-S0 galaxies with old stellar population for the PTF sample might be due to that the birth rate at redshift $\sim$ 0.1 is significantly lower than that at local universe. We further investigate the host stellar mass and present the cumulative fractions for host metallicity and stellar mass distribution of the two samples in Figs.~\ref{metallicity} and ~\ref{mass}. 
    
The PTF SN Ia sample at an average redshift of 0.073 shows an obvious preference to reside in metal-poor and less massive galaxies compared to our local sample of SNe Ia. K-S tests give p values of 0.167 and 0.073 for the host metallicity and stellar mass distribution of the two samples, respectively. This further confirms a metal-poor environment for the SN Ia population at a redshift $\sim$ 0.1 compared to the local population. In Fig. 1 of \citet{Briel2022}, we could identify the effect of metallicity on the DTD of SNe Ia, that is, the fraction of SNe Ia with long delay times (i.e., $>$ 10 Gyr) is significantly higher for metal-rich populations. Thus, the group of SNe Ia with long delay times should contribute more to the local SN Ia rate, which is consistent with our findings.
Also, the host stellar mass of the PTF sample is significantly lower than that of our sample, considering the lack of E-S0 host galaxies for the PTF sample and the fact that E-S0 galaxies are, on average, massive evolved galaxies with mostly old stellar populations. We conclude that the higher SN Ia rate at z $\sim$ 0 originates mainly from massive E-S0 galaxies.  
Meanwhile, the stellar population of E-S0 galaxies is on average older than late-type spirals, so the SNe Ia exploding in E-S0 galaxies contributes mostly to the delayed component with long delay times. In Sect.~\ref{DTD} we have shown that the high SN Ia rate in the local universe is mainly due to the contribution of a Gaussian component of DTD centered at a delay time of $\sim$ 12.5 Gyr. This is perfectly consistent with the fact that the number of SNe Ia exploding in metal-rich and massive E-S0 galaxies is greater at z $\sim$ 0 than the PTF sample at z = 0.073.

\section{Conclusions} \label{sum}
	
In this paper, we present the local volumetric rates for type Ia and core-collapse supernovae. 
	
We used the SN sample constructed in Paper I and performed a Monte Carlo simulation to estimate the local volumetric rate for each type of SNe. Uncertainties involved with the Monte Carlo simulation are also estimated.
We find the SN Ia rate in the local universe to be 
	$$\mathrm{SNR_{Ia} = 0.325\pm0.040^{+0.016}_{-0.010} \times 10^{-4} yr^{-1} Mpc^{-3} h^3_{70}},$$
	and the local volumetric rate for core-collapse supernovae is
	$$\mathrm{SNR_{CC} = 0.688\pm0.078^{+0.206}_{-0.027} \times 10^{-4} yr^{-1} Mpc^{-3} h^3_{70}}.$$
The results are consistent with recent rate measurements, although the SN Ia rate is relatively larger. We achieved significantly better precision than most previous studies, similar to the precision of the PTF and ASAS-SN survey.

Combined with the GLADE+ sample, we also calculated the supernova rate as a function of galaxy Hubble type. The SNuM rates of SNe Ia, Ibc and II for galaxies with fiducial size are generally consistent with \citet{Li2011b}, except for the noticeable higher SN Ia rate in Sc galaxies. This excess is consistent with the number distribution discussed in Paper I, but the SNuM rate of SN Ia in E-S0 galaxies does not show a peak structure as in the number distribution. 
The significantly small SN Ibc rate in Scd and Irr galaxies compared to \citet{Li2011b} might be due to the small sample size for SNe Ibc or that \citet{Li2011b} overestimated the rate for their preference to observe massive Scd and Irr galaxies and thus the absence of abundant low mass galaxies. We also estimated the Galactic SN rate to be $3.08 \pm 1.29$ SNe per century, which is in good agreement with historical results of 1.4-5.8 SNe per century.

Finally, we combined our result with other literature sample of SN rates up to higher redshifts. We used the cosmic SN Ia rate evolution to constrain power-law, e-folding and two-component delay-time distribution models. The DTD models were convolved with three different SFHs: the \citet{Yuksel2008}, \citet{Li2008}, and \citet{Harikane2022} SFH model. Power-law and e-folding DTD models could fit the overall evolution of the rate but failed in the local universe to predict a nearly constant rate rather than a significant decline from z = 0 to $\sim$ 0.1. The two-component DTD model provided a proper fit to the rate evolution with similar values of $\chi^2$. More importantly, it was able to fit the SN Ia rate decline in the local universe and gave one prompt component following the power-law distribution and a delayed Gaussian component centered at a delay time of 12.63 $\pm$ 0.38 Gyr, suggesting the existence of star-forming galaxies at $z > 6$. This two-component model is consistent with the double peak structure we identified in the Hubble type distribution of SNe Ia in Paper I. The long delay time of the delayed component favors the DD channel according to previous model simulations, and more recent studies further confirmed the existence of the group of SNe Ia with extremely long delay times. By comparing the host environment of our SN Ia sample with that of the PTF sample at an average redshift of 0.073, we found that the higher SN Ia rate in the local universe compared to z $\sim$ 0.1 mainly comes from SNe Ia exploding in massive E-S0 galaxies with old stellar populations. This further implies the existence of a group of SN Ia with very large delay times (i.e., the Gaussian component in our two-component model). Our local rate of CCSNe is perfectly consistent with the cosmic SFHs from the literature, as predicted by theory.
	
From our study, we find that the nearby SNe discovered in recent years suffers from a severe bias beyond 40 Mpc. Only the sample inside this distance seems to be relatively complete. The further development of rolling search surveys could help better constrain the measurement of the rate beyond z = 0.01 and improve our understanding of the DTD and progenitor systems of SNe Ia.
	
\begin{acknowledgements}	
We acknowledge the support of the staff of Lijiang 2.4m and Xinglong 2.16-m telescopes. This work is supported by the National Science Foundation of China (NSFC grants 12288102, 12033003, and 11633002), the science research grant from the China-Manned Space Project No. CMS-CSST-2021-A12, and the Tencent Xplorer prize. This work makes use of observations from the Las Cumbres Observatory global telescope network. The LCO group is supported by NSF grants AST-1911225 and AST-1911151. DDL is supported by the National Key R\&D Program of China (Nos. 2021YFA1600403 and 2021YFA1600400), the National Natural Science Foundation of China (No. 12273105, 12288102), the Youth Innovation Promotion Association CAS (No. 2021058), the Yunnan Revitalization Talent Support Program – Young Talent project, and the Yunnan Fundamental Research Projects (Nos 202401AV070006 and 202201AW070011). We acknowledge the usage of the HyperLeda database (http://leda.univ-lyon1.fr). H.Lin is supported by the National Natural Science Foundation of China (NSFC, Grant No. 12403061) and the innovative project of "Caiyun Post-doctoral Project" of Yunnan Province.
\end{acknowledgements}	
	
\section*{Data Availability}
	
The data underlying this article will be shared on reasonable request to the corresponding author.

\bibliographystyle{aa}
\bibliography{reference}

\begin{appendix}

    \onecolumn

    \section{Detection efficiency of the three surveys.} 

    In this section, we present the detection efficiencies for the three surveys, (i.e., ZTF, ATLAS, and ASAS-SN), adopted in this work.
    
\begin{figure*}[h!]
    \centering
		\includegraphics[width=0.6\textwidth]{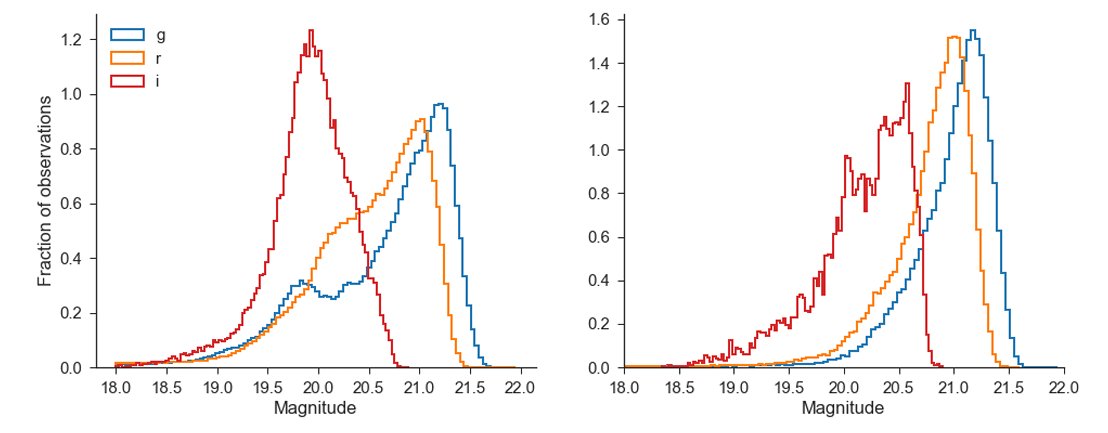}
		\caption{Left: Histogram of 5$\sigma$ limiting magnitudes in 30 second exposures for g (blue), r (orange), and i (red) bands over one lunation. Right: Limiting magnitudes for observations obtained within 3 days of new moon. This is Figure. 6 from \citet{Bellm2019}.}\label{ZTF}
    \end{figure*}

    In Fig.~\ref{ZTF} we present the histograms of 5$\sigma$ limiting magnitudes of ZTF, the figure is adopted from \citet{Bellm2019}. Median 5$\sigma$ limiting magnitudes are 20.8 mag in g-band, 20.6 mag in r-band, and 19.9 in i-band.
   
\begin{figure*}[h!]
    \centering
		\includegraphics[width=0.6\textwidth]{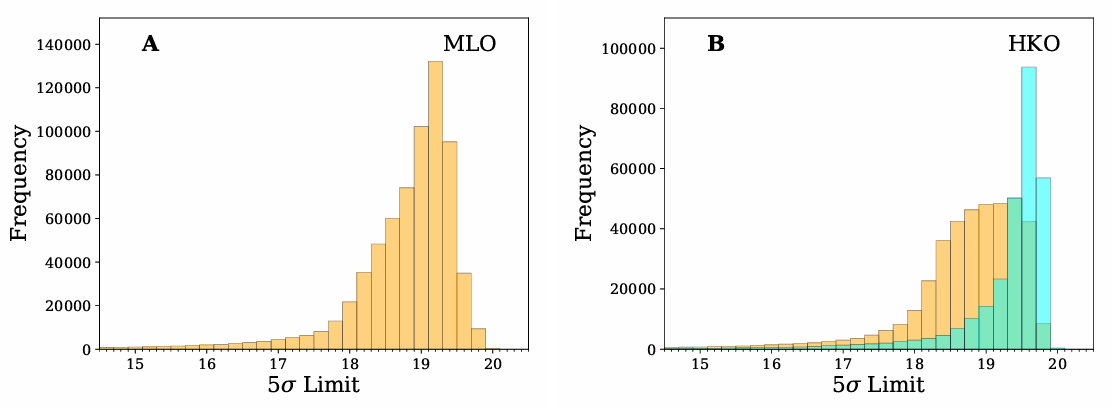}
		\caption{A:Histogram of 5$\sigma$ limiting magnitudes for MLO in the $o$ band.B:Same as in A,but for HKO with the c and o bands plotted together. This is Figure. 8 of \citet{Smith2020}.}\label{ATLAS}
    \end{figure*}

    In Fig.~\ref{ATLAS} we present the histograms of 5$\sigma$ limiting magnitudes of ATLAS, the figure is adopted from \citet{Smith2020}. Median 5$\sigma$ limiting magnitudes for each of the systems are 19.0 mag (ATLAS-HKO o band), 19.6 mag (ATLAS-HKO c band), and 19.0 mag (ATLAS MLO o band). With these distributions we compare the apparent magnitude of an SN to calculate the probability of this SN being detected at $>$ 5$\sigma$ above the background noise by these two surveys and hence the probability that the 5$\sigma$ limiting magnitude is dimmer than the SN.

\begin{figure}[h!]
        \centering
		\includegraphics[width=0.35\columnwidth]{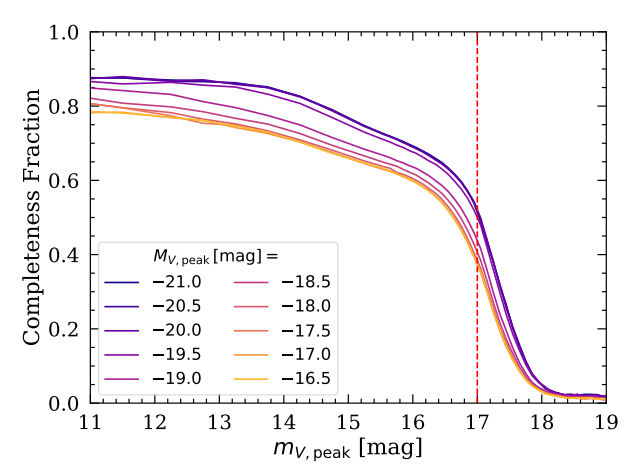}
		\caption{Completeness as a function of peak apparent magnitude $\mathrm{m_{V,peak}}$ with the peak absolute magnitudes $\mathrm{M_{V,peak}}$ shown in different colors. The vertical dashed red line marks our standard choice for the limiting magnitude, $\mathrm{m_{V,lim}}$ = 17 mag, where the completeness is $\sim$ 50\%. This is Figure. 4 from \citet{Desai2024}.}\label{ASASSN}
    \end{figure}

    In Fig.~\ref{ASASSN} we present the detection completeness as a function of peak apparent magnitude $\mathrm{m_{V,peak}}$ for ASAS-SN, the figure is adopted from \citet{Desai2024}. The standard choice for the limiting magnitude is $\mathrm{m_{V,lim}}$ = 17 mag, where the completeness is $\sim$ 50\%. For SN with peak apparent magnitude outside the values given in Fig.~\ref{ASASSN}, we did a linear interpolation to determine the completeness curve.

    \twocolumn

    \section{Peak absolute magnitude distributions of different sub-types.} 
    
    \begin{table}[h!]
		\centering
		\caption{Bias-corrected B-band peak absolute magnitude distributions of different sub-types given by \citet{Richardson2014}.}\label{MB}
		\begin{threeparttable}
			
			\begin{tabular}{ccc}
				\toprule
				Type & $\overline{M_{B}}$ & $\sigma$\tablefootmark{a}  \\
				\midrule
				Ia & $-19.25\pm0.20$ & 0.50 	\\
				Ib & $-17.45\pm0.33$ & 1.12 	\\
				Ic & $-17.66\pm0.40$ & 1.18 	\\
                IIP & $-16.75\pm0.37$ & 0.98 	\\
                IIL & $-17.98\pm0.34$ & 0.86 	\\
                IIb & $-16.99\pm0.45$ & 0.92 	\\
                IIn & $-18.53\pm0.32$ & 1.36 	\\
				\bottomrule
			\end{tabular}
			
		\end{threeparttable}

        \tablefoot{\tablefoottext{a}{The statistical standard deviation in the mean.}}
        
    \end{table}
    
    \section{Summary of SN rate measurements}

     \begin{table}[h!]
		\centering
		\caption{All CCSN rate measurements used in this work.}\label{IIrateall}
		\begin{threeparttable}
		 
			\begin{tabular}{ccc}
				\toprule
				z & Rate$^{a}/10^{-4} \mathrm{yr^{-1} Mpc^{-3} h^3_{70}}$ & Reference  \\
				\midrule
                0.1-0.5 & $2.51^{+0.88+0.75}_{-0.75-1.86}$ & (6) \\
                0.5-0.9 & $3.96^{+1.03+1.92}_{-1.06-2.60}$ & (6) \\
                0.26 & $1.789^{+0.65}_{-0.57}$ & (4) \\
                0.21 & $1.15^{+0.43+0.42}_{-0.33-0.36}$ & (2) \\
                0.3 & $1.63 \pm 0.30^{+0.32}_{-0.24}$ & (1) \\
			  $\sim 0$ & $0.705 \pm 0.211$ & (10) \\
				0.003 & $1.626 \pm 0.407$ & (3) \\
                0.39 & $3.00^{+1.28+1.04}_{-0.94-0.57}$ & (7) \\
                0.73 & $7.39^{+1.86+3.20}_{-1.52-1.60}$ & (7) \\
                1.11 & $9.57^{+3.76+4.96}_{-2.80-2.80}$ & (7) \\
                0.39 & $3.29^{+3.08+1.98}_{-1.78-1.45}$ & (11) \\
                0.73 & $6.40^{+5.30+3.65}_{-3.12-2.11}$ & (11) \\
				0.072 & $1.06 \pm 0.11 \pm 0.15$ & (13) 	\\
                0.075 & $1.04^{+0.33+0.04}_{-0.26-0.11}$ & (9) \\
                0.05-0.15 & $1.13^{+0.62+0.49}_{-0.53-0.49}$ & (5)	\\
                0.15-0.35 & $1.21 \pm 0.27 \pm 0.47$ & (5) 	\\
                0.1-0.5 & $1.97^{+1.45}_{-0.85}$ & (12) 	\\
                0.5-0.9 & $2.68^{+1.54}_{-1.04}$ & (12) 	\\
                0.9-1.3 & $1.70^{+1.19}_{-0.71}$ & (12) 	\\
                1.3-1.7 & $3.25^{+2.03}_{-1.32}$ & (12)	\\
                1.7-2.1 & $3.16^{+3.37}_{-1.77}$ & (12) 	\\
                2.1-2.5 & $6.17^{+6.76}_{-3.52}$ & (12)	\\
                0.028 & $0.910^{+0.156}_{-0.127}$ & (8)	\\
                0-0.01 &  $0.688 \pm 0.078^{+0.206}_{-0.027}$ & this work	\\
				\bottomrule
			\end{tabular}
          
		\end{threeparttable}

        \tablefoot{\tablefoottext{a}{Uncertainty is split between statistical and systematic uncertainty.}}

        \tablebib{(1) \citet{Bazin2009}; (2) \citet{Botticella2008}; (3) \citet{Botticella2012}; (4) \citet{Cappellaro2005}; (5) \citet{Cappellaro2015}; (6) \citet{Dahlen2004}; (7) \citet{Dahlen2012}; (8) \citet{Frohmaier2021}; (9) \citet{Graur2015}; (10) \citet{Li2011b}; (11) \citet{Melinder2012}; (12) \citet{Strolger2015}; (13) \citet{Taylor2014}.}
        
    \end{table}
    
    \begin{table}[h!]
		\centering
		\caption{All SN Ia rate measurements used in this work.}\label{Iarateall}
		\begin{threeparttable}
		 
			\begin{tabular}{ccc}
				\toprule
				z & Rate\tablefootmark{a}$/10^{-4} \mathrm{yr^{-1} Mpc^{-3} h^3_{70}}$ & Reference  \\
				\midrule
				$\sim$ 0  & $0.185 \pm 0.05$ & (4)  \\
                0.14  & $0.28^{+0.22+0.07}_{-0.13-0.04}$ & (11) \\
				0.098  & $0.24 \pm 0.12$ & (14) \\
                0.13   & $0.158^{+0.056}_{-0.043} \pm 0.035$ & (2) \\
                0.2-0.6   & $0.69^{+0.34+1.54}_{-0.27-0.25}$ & (6) \\
                0.6-1.0   & $1.57^{+0.044+0.75}_{-0.25-0.53}$ & (6) \\
                1.0-1.4   & $1.15^{+0.47+0.32}_{-0.26-0.44}$ & (6) \\
                1.4-1.8   & $0.44^{+0.32+0.14}_{-0.25-0.11}$ & (6) \\
                0.25  & $0.17 \pm 0.17$ & (1)  \\
                0.3  & $0.34^{+0.16+0.21}_{-0.15-0.22}$ & (3) \\
				0.09 & $0.29^{+0.09}_{-0.07}$ & (8)  	\\
                0.2 &   $0.189^{+0.042+0.018}_{-0.034-0.015}$ & (12)  	\\
                0.025-0.050  & $0.278^{+0.112+0.015}_{-0.083-0.000}$ & (9)\\
                0.075-0.125  & $0.259^{+0.052+0.018}_{-0.044-0.001}$ & (9)\\
                0.125-0.175  & $0.307^{+0.038+0.035}_{-0.034-0.005}$ & (9)\\
                0.175-0.225  & $0.348^{+0.032+0.082}_{-0.030-0.007}$ & (9)\\
                0.225-0.275  & $0.365^{+0.031+0.182}_{-0.028-0.012}$ & (9)\\
                0.275-0.325  & $0.434^{+0.037+0.396}_{-0.034-0.016}$ & (9)\\
                $\sim 0$  & $0.301 \pm 0.062$ & (13)  	\\
                0.1-0.2  & $0.14 \pm 0.09^{+0.06}_{-0.12}$ & (16)\\
                0.2-0.3  & $0.28 \pm 0.07^{+0.06}_{-0.07}$ & (16)\\
                0.3-0.4  & $0.36 \pm 0.06^{+0.05}_{-0.06}$ & (16)\\
                0.4-0.5  & $0.36 \pm 0.06^{+0.04}_{-0.05}$ & (16)\\
                0.5-0.6  & $0.48 \pm 0.06^{+0.04}_{-0.05}$ & (16)\\
                0.6-0.7  & $0.48 \pm 0.05^{+0.04}_{-0.06}$ & (16)\\
                0.7-0.8  & $0.58 \pm 0.06^{+0.05}_{-0.07}$ & (16)\\
                0.8-0.9  & $0.57 \pm 0.05^{+0.06}_{-0.07}$ & (16)\\
                0.9-1.0  & $0.77 \pm 0.08^{+0.10}_{-0.12}$ & (16)\\
                1.0-1.1  & $0.74 \pm 0.12^{+0.10}_{-0.13}$ & (16)\\
                0.25 &   $0.36^{+0.60+0.12}_{-0.26-0.35}$ & (17)  	\\
                0.75 &   $0.51^{+0.27+0.23}_{-0.19-0.19}$ & (17) 	\\
                1.25 &   $0.64^{+0.31+0.34}_{-0.22-0.23}$ & (17) 	\\
                1.75 &   $0.72^{+0.45+0.50}_{-0.30-0.28}$ & (17)  	\\
                2.25 &   $0.49^{+0.95+0.45}_{-0.38-0.24}$ & (17)  	\\
                0.05-0.15  & $0.55^{+0.50}_{-0.29} \pm 0.20$ & (5) 	\\
                0.15-0.35   & $0.39^{+0.13}_{-0.12} \pm 0.10$ & (5) 	\\
                0.35-0.55   & $0.52^{+0.11}_{-0.13} \pm 0.16$ & (5) 	\\
                0.55-0.75   & $0.69^{+0.19}_{-0.18} \pm 0.27$ & (5) 	\\
                0.073  & $0.243^{+0.029+0.033}_{-0.029-0.019}$ & (10)	\\
                < 0.1  & $0.235 \pm 0.024$ & (15) 	\\
                0.01-0.04  & $0.291^{+0.058}_{-0.045}$ & (18)	\\
                0.024  & $0.228 \pm 0.020$ & (7) \\
                0-0.01 & $0.325 \pm 0.040^{+0.016}_{-0.010}$ & this work 	\\
				\bottomrule
			\end{tabular}
            
		\end{threeparttable}

        \tablefoot{\tablefoottext{a}{Uncertainty is split between statistical and systematic uncertainty.}}

        \tablebib{(1) \citet{Barris2006}; (2) \citet{Blanc2004}; (3) \citet{Botticella2008}; (4) \citet{Cappellaro1999}; (5) \citet{Cappellaro2015}; (6) \citet{Dahlen2004}; (7) \citet{Desai2024}; (8) \citet{Dilday2008}; (9) \citet{Dilday2010}; (10) \citet{Frohmaier2019}; (11) \citet{Hardin2000}; (12) \citet{Horesh2008}; (13) \citet{Li2011b}; (14) \citet{Madgwick2003}; (15) \citet{Perley2020}; (16) \citet{Perrett2012}; (17) \citet{Rodney2014}; (18) \citet{Sharon2022}.}
        
    \end{table}

\FloatBarrier

\end{appendix}

\end{document}